\def\hii{{\rm H\,{\sc ii} }}
\def\ha{{\rm H$\alpha$ }}

\documentstyle[draft,epsf,psfig]{mn}

\title{Exploring the evolution of spiral galaxies}

\author[Bell \& Bower]
{
Eric F. Bell$^{1,2,3}$ and
Richard G. Bower$^{1,4}$\\
$^1$ Department of Physics, University of Durham, 
Science Laboratories, South Road, 
Durham DH1 3LE, UK\\
$^2$ Steward Observatory, University of Arizona, 
933 N. Cherry Ave., Tucson, AZ 85721, USA\\
$^3$ ebell@as.arizona.edu\\
$^4$ r.g.bower@durham.ac.uk}
\begin{document}

\date{\fbox{{\sc Re-submitted to MNRAS 12 May 2000} }}

\maketitle

\begin{abstract}
We have constructed a family of simple models for spiral galaxy evolution
to allow us to investigate observational
trends in star formation history with galaxy parameters.  
The models are used to generate broad band colours from which 
ages and metallicities are derived in the same way as the data.  
We generate a grid
of model galaxies and select only those which lie
in regions of parameter space covered by the sample.
The data are consistent with the proposition that the star formation 
history of a region within a galaxy depends primarily on the 
local surface density of the gas but that one or two additional
ingredients are required to fully explain the observational data.  
The observed age gradients appear
steeper than those produced by the density dependent star formation law, 
indicating that the star formation law or infall
history must vary with galactocentric radius.  
Furthermore, the metallicity--magnitude and
age--magnitude correlations are not reproduced by a local
density dependence alone. These correlations
require one or both of the following: 
(i) a combination of mass dependent infall {\it and}
metal enriched outflow, or (ii) a mass dependent galaxy formation epoch.
Distinguishing these possibilities on the basis of current data is
extremely difficult.
\end{abstract}


\section{Introduction} \label{c4intro}

The chemical evolutionary histories of spiral
galaxies provide considerable insight into 
many of the important processes involved in galaxy formation 
and evolution.  For example, we can study
star formation laws (SFLs; e.g.\ Wyse \& Silk 1989; Phillipps \& Edmunds 1991),
the interactions between newly-formed stars and the interstellar 
medium (e.g.\ Dekel \& Silk 1986; MacLow \& Ferrera 1999), 
the importance and effects of gas flows (e.g.\ Lacey \& Fall
1985; Edmunds 1990; Edmunds \& Greenhow 1995) and the infall
that must accompany disc formation (e.g.\ Tinsley \& 
Larson 1978; Lacey \& Fall 1983; Steinmetz \& M\"{u}ller 1994).  
The main challenge is obtaining
{\it unambiguous} insight into particular physical processes. 
Some of the ambiguity can be circumvented by 
studying both the ages and the metallicities of galaxies: in this paper
we use the ages and metallicities of a sample of face-on spiral galaxies
to constrain which processes are the most important in 
affecting their observational properties.

Despite these difficulties, considerable 
progress has been made in understanding 
some important aspects of galaxy formation and evolution.  
A local density dependence in the SFL is 
strongly favoured, although other factors may affect the star 
formation rate (SFR) over galactic scales 
(e.g.\ Schmidt 1959; Dopita 1985; Kennicutt 1989; Wyse \& Silk 1989; 
Dopita \& Ryder 1994; Prantzos \& Aubert 1995; Kennicutt 1998).
Infall may be important in determining 
the metallicity distribution of stars in the solar neighbourhood
(e.g.\ Tinsley 1980; Prantzos \& Aubert 1995; Pagel 1998).
Other processes are more controversial: e.g.\
metal-enriched outflows (e.g. MacLow \& Ferrera 1999)
or radial gas flows (e.g.\ Edmunds \&
Greenhow 1995; Lacey \& Fall 1985).

However, recent observational advances, coupled with the 
development of multiple metallicity
stellar population synthesis codes has allowed comparison of
galaxy evolution models
with both the gas metallicities and colours of spiral galaxies (Contardo, 
Steinmetz \& Fritze-von Alvensleben 1998; Jimenez et al.\ 1998; 
Boissier \& Prantzos 2000;
Prantzos \& Boissier 2000; Cole et al.\ 2000).  The colours of spiral
galaxies depend on both their ages and metallicities, therefore study of 
their colours offers fresh insight into galaxy formation and evolution, 
although inevitably degeneracies remain.

In Bell \& de Jong (2000; BdJ hereafter), we analysed the
optical--near-infrared (near-IR) colours of a sample
of 121 low-inclination spiral galaxies in conjunction with up-to-date
stellar population synthesis models to explore trends in age and metallicity
with galaxy parameters, such as magnitude or surface brightness.
In particular, we found that there are significant trends between the age
and $K$ band surface brightness of a galaxy, and between the 
metallicity and both the $K$ band magnitude and surface brightness of 
a galaxy.  In that paper, we argued that these correlations could be the 
result of a surface density-dependent SFL, coupled
with galaxy mass-dependent chemically-enriched gas outflows.  

In this paper, we investigate these ideas in more detail. We use
a family of simple
models to explore the effects of infall, outflows, 
age differences and SFLs on
the colour-based ages and metallicities of spiral galaxies.  Our aim
is not to construct a self-consistent, realistic model of galaxy 
formation and evolution.
This work is intended to guide future, more detailed explorations
of the star formation histories (SFHs) of spiral galaxies:  this
simple modelling isolates which physical processes affect which 
observables, to allow more realistic models to concentrate on formulating
self-consistent prescriptions for the most important physical phenomena.

The plan of this paper is as follows.
In section \ref{c4model}, we outline the data and its main limitations.  
We describe the chemical evolution 
model, the basic assumptions and equations and outline
how we translate the model output into observables which we can 
readily compare with the data.  In section \ref{c4evo}, we 
describe the properties of the closed box model.
In section \ref{c4inf} we explore
the effects of infall, outflow and systematic trends in 
galaxy formation epoch.  In section 
\ref{c4sfl}, we investigate the effects of changing the SFL on 
our results.  In section \ref{c4disc} we discuss the results
further, checking the plausibility of these models with other observational
constraints.  There, we also compare our models to a comparable, but
more detailed model by Boissier \& Prantzos.
Finally, in section \ref{c4conc}, we summarise our results.

\section{The Method} \label{c4model}

\subsection{The data}

For this paper, we use the sample of 121 low-inclination spiral 
galaxies from BdJ.    The sample is described in more 
detail in BdJ and in the sample's source papers 
\cite{papi,dji,tv1}.  
The sample galaxies were chosen to have radially-resolved
surface brightnesses in at least one near-IR and two optical passbands.
The sample galaxies have a wide range of surface brightnesses, magnitudes, scale lengths
and gas fractions, but are not complete in any statistical sense (at least
as a unit).

In BdJ, we used a combination of at least
one near-IR and two optical passbands to split (to some degree)
the age-metallicity degeneracy.  We fit simplified SFHs to 
the optical--near-IR colours using a maximum likelihood technique
to derive crude age (reflecting
the amount of recent to past star formation; c.f. a birthrate parameter)
and metallicity estimates.  These estimates are not
accurate in an absolute sense: they are subject to uncertainties
from a number of sources including model uncertainties, the effects
of small bursts of star formation, the assumption of a single
epoch of galaxy formation 12 Gyr ago and dust, to name a few.  However,
we argue that the estimates are robust in a relative sense:
these uncertainties compromise the absolute ages and metallicities but
leave the relative ranking of galaxies by age or metallicity relatively
unaffected.  

The ages and metallicities of one scale length wide annuli in our
sample galaxies were estimated using the above method.
We also constructed estimates of the age and metallicity at one disc
half-light radius and their gradients per $K$ band scale length using
a weighted linear fit to the ages and metallicities as a function of 
radius.  More description of the method, its caveats and limitations
can be found in BdJ.

\subsection{Basic assumptions and equations}

In order to make the investigation of the trends in age and 
metallicity a tractable problem, we adopt
highly simplified {\it ad hoc} prescriptions describing 
star formation and galaxy evolution.  These simple approximations allow us 
to investigate which effects play an important 
role in e.g.\ imprinting mass dependence in the
SFH.  We {\it do not} include gas flows in these models:
assuming that the final total baryonic mass distribution is
no different from a model without gas flows, the primary difference
between models with and without gas flows will be the metallicity
gradients (e.g.\ Lacey \& Fall 1985; Edmunds 1990; Edmunds \&
Greenhow 1995).  Therefore, the metallicity
gradients are not iron-clad constraints on the models (in 
any case, the metallicity gradients are relatively
unaffected by many of the changes explored in this paper,
so the metallicity gradients were not particularly strong
model constraints anyway).

For consistency with our age and metallicity estimation procedure,
we assume that our model galaxy forms 12 Gyr ago as an 
exponential disc of gas with surface density 
$\Sigma_0(r) = \Sigma_0(r=0) e^{-r}$, where $\Sigma_0$ is 
the initial surface density of gas in M$_{\sun}$\,pc$^{-2}$, 
and $r$ is the radius in units of the scale length of 
the gas (denoted by $h$).   For the infall case (see section \ref{c4infall}),
$\Sigma_0(r) = 0$ initially, and the gas mass is gradually built up over
time assuming an exponentially declining infall rate 
with $e$-folding time $\tau_{\rm infall}$.
For the case in which we allow the galaxy
formation epoch to vary as a function of 
its mass (see section \ref{c4agediff}),
we change the galaxy formation epoch 
from 12 Gyr to between 4 and 12 Gyr, depending on galaxy mass.

The gas forms stars according to a prescribed SFL: in much of this 
paper we adopt a Schmidt (1959) SFL in terms of the gas 
surface density $\Sigma_{\rm gas}$:
\begin{equation}
\psi = k\Sigma_{\rm gas}^n,
\end{equation}
where $\psi$ is the SFR in M$_{\sun}$\,pc$^{-2}$\,Gyr$^{-1}$,
$k$ is the efficiency of star formation at a gas surface density of 
1 M$_{\sun}$\,pc$^{-2}$ and $n$ is the exponent specifying how sensitively 
the SFR depends on gas surface density.

This star formation produces heavy elements; here we adopt the instantaneous
recycling approximation (IRA; e.g.\ Tinsley 1980; Pagel 1998).  A fraction
$R$ of the mass of newly formed stars is instantaneously returned to 
the gas.  We adopt a Salpeter (1955) initial mass function (IMF) 
with lower and upper
mass limits of 0.1 M$_{\odot}$ and 125 M$_{\odot}$ respectively
to describe the chemical and photometric evolution of our stellar
populations: for this IMF the returned fraction $R \sim 0.3$.
This gas is returned along with a mass $p\psi(1-R)$ of heavy elements, where
$p$ is the true yield, and is defined as the mass of freshly produced 
heavy elements per unit mass locked up in long-lived stars.  
The true yield $p$ is taken to 
be 0.02 (solar metallicity) hereafter unless explicitly stated otherwise.  
Note that, for simplicity, 
we assume a metallicity-independent yield.  
Note that our use of the IRA should not lead to significant inaccuracies, 
as the metallicity of spiral galaxies is typically measured via their 
oxygen content:  for oxygen, the IRA
is a fairly accurate approximation as it is produced primarily
by Type II supernovae (c.f.\ Pagel 1998; although this assumption
breaks down at late stages of galactic evolution near gas exhaustion; 
Portinari \& Chiosi 1999; Prantzos \& Boissier 2000).

Once the IRA is adopted, the following three equations specify the 
evolution of the galaxy completely:
\begin{eqnarray}
\frac{d\Sigma_{\rm gas}}{dt} & = & F - E - \psi(1-R) \\
\frac{d\Sigma_{\rm stars}}{dt} & = & \psi(1-R) \\
\frac{d(\Sigma_{\rm gas}Z)}{dt} & = & p\psi(1-R) - Z\psi(1-R) 
	- {Z_E}E + {Z_F}F, 
\end{eqnarray}
where $\Sigma_{\rm stars}$ is the surface density of the stars in 
M$_{\sun}$\,pc$^{-2}$, $F$ is the gas surface density infall rate (with 
an initial metallicity $Z_F$; we assume $Z_F = 0$ hereafter)
in M$_{\sun}$\,pc$^{-2}$\,Gyr$^{-1}$, $E$ is the surface density 
of gas ejected in outflows (with metallicity $Z_E$) in  
M$_{\sun}$\,pc$^{-2}$\,Gyr$^{-1}$ and $Z$ is the gas metallicity
(Tinsley 1980; Pagel 1998).

\subsection{Determining ages and metallicities} \label{subsec:c4det}

We follow the evolution of the galaxy using a numerical 
scheme with a 20 Myr timestep.  We split our model galaxies 
into 20 radial zones between $r = 0$ and $r = 4$ gas disc scale
lengths, to allow study of both global and radial 
trends in age and metallicity.  While some of the cases we study here 
have analytical solutions, our use of a numerical scheme allows us to use
more complex SFLs and e.g.\ infall or outflow histories.

In order to properly
compare the models with the data, we use the colour-based 
maximum-likelihood technique from BdJ 
to determine the ages and metallicities of our model galaxies.
In order to use this technique, we must have a set of optical
and near-IR colours for our model galaxies.
Therefore, in each zone at each timestep,
we use the {\it total} mass of newly-formed stars (both short- and 
long-lived) to compute the contribution of those 
stars to the total flux at the present day in 
$U$, $B$, $V$, $R$, $I$, $J$, $H$ and $K$ bands using interpolations
between the multi-metallicity {\sc gissel98} models of Bruzual 
\& Charlot (in preparation).  

We then use these local colours
as input to the maximum-likelihood age and metallicity estimator
developed for and presented in BdJ.  In this way, 
{\it we obtain model ages and metallicities determined
in exactly the same way as the observations we compare with}.  This 
can be quite important:  especially so for older stellar populations.
Mass-weighted average ages of older stellar populations
can differ considerably from the luminosity-weighted 
ages derived using the colour-based technique 
because of only relatively modest
amounts of recent star formation.   

Using these local age and metallicity estimates, we construct 
estimates of the global age and metallicity gradients and intercepts
(at the $K$ band disc half-light radius), 
using an unweighted least-squares fit (c.f.\  
BdJ).  $K$ band disc central 
surface brightnesses and scale lengths are determined by fitting the first 
3 (gas) disc scale lengths of the surface brightness profile.  Global 
gas fractions are determined by direct summation of the model gas and 
stellar masses.

The sample galaxies from BdJ cover a broad
range of magnitudes and surface brightnesses.  Therefore, 
to provide a fair comparison, the model galaxies must cover
a similarly broad range of magnitudes and surface brightnesses.
We adopt an empirical approach: we run a grid of 357 models 
with total (baryonic and dark) masses between $10^9$ M$_{\sun}$ and 
$10^{14}$ M$_{\sun}$ (we assume a baryon fraction of 0.05 hereafter), 
and central baryonic surface densities between 
$10^{0.5}$ M$_{\sun}$\,pc$^{-2}$ and $10^{4.5}$ M$_{\sun}$\,pc$^{-2}$.
The step size is 0.25 dex.  This range is sufficient to cover the 
full observed range of parameter space probed in BdJ,
assuming that the baryonic content of a galaxy turns
entirely into solar-type stars.  

\begin{figure}
\begin{center}
\psfig{figure=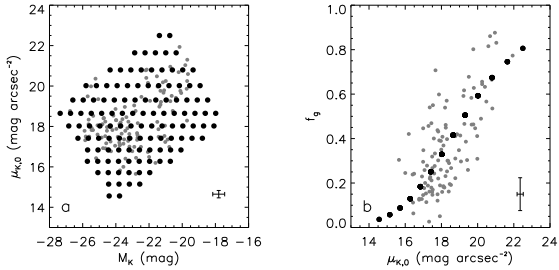,width=8cm}
\end{center}
\caption{{\bf Fiducial Model: }
Correlation between $K$ magnitudes and   
$K$ band central surface brightnesses (panel a),
and between $K$ band central
surface brightnesses 
and gas fractions (panel b).
The data are from BdJ and are shown as grey circles (average error
bars are in the bottom right corner of each plot)
and the fiducial model is shown by black circles. 
Panel a shows our selection criteria based on 
the observed distribution of galaxies in the $M_K$ vs.\ $\mu_{K,0}$ plane.
Panel b shows the surface brightness--gas fraction
correlation, which we use as a constraint when searching for 
adequate model fits to the data. 
}
\label{fig:c4phys}
\end{figure}

However, this situation is complicated by the (broad) correlation between 
surface brightness and magnitude (c.f.\ Fig.\ 12 from 
BdJ, or the grey dots in panel a of Fig.\ \ref{fig:c4phys}).  
Because of this correlation between 
surface brightness and magnitude, any correlation between e.g.\ 
age and surface brightness will automatically 
translate into a correlation between 
age and magnitude.  However, our model grid does not incorporate
this correlation.  Therefore, to provide a fair comparison with the 
data, we select galaxies from the model
grid that fall within the region inhabited by the sample galaxies in the 
$K$ band surface brightness--absolute magnitude plane (Fig.\ 
\ref{fig:c4phys}) using the following criteria:
\begin{eqnarray}
\mu_{K,0} & > & 13 + 0.6(M_K + 26) \\
\mu_{K,0} & < & 20 + 0.6(M_K + 26) \\
\mu_{K,0} & > & 19 - 1.25(M_K + 28) \\
\mu_{K,0} & < & 23 - 1.25(M_K + 21), 
\end{eqnarray}
where $\mu_{K,0}$ is the $K$ band disc 
central surface brightness and $M_K$ is the 
$K$ band absolute magnitude of the galaxy.  In this way, we can 
empirically select galaxies with a range of physical parameters
consistent with those taken from BdJ.
This approach ensures that the galaxies we produce automatically
roughly satisfy the selection criteria of BdJ's sample.
We make no attempt to derive the allowed range of 
surface brightnesses and scale lengths on the basis of the
angular momentum of infalling gas (e.g.\
Dalcanton, Spergel \& Summers 1997; Mo, Mao \& White 1998).

Note that galaxies with different sets of surface densities and 
masses may be chosen, depending on the model details (especially on 
the efficiency with which stars are turned into gas). 
We show an example of the selection box in panel a of 
Fig.\ \ref{fig:c4phys} for the fiducial closed box model 
discussed in the next section.
We show the $K$ band central surface brightness against the $K$ band absolute
magnitude of the data (in grey) and the fiducial model (in black).
The model points show clearly the selection criteria that are 
applied on the model central surface brightnesses and 
magnitudes: these selection limits are applied to preserve the 
broad correlation between surface brightness and magnitude
in the dataset, and to make sure we do not compare
the data with models of galaxies that are drastically
different from those in the data.
Panel b of Fig.\ \ref{fig:c4phys} shows the $K$ band
central surface brightness against gas fraction relation for
the same fiducial model.  All of the
models presented in this paper are constrained to 
reproduce this correlation (although, in the case of e.g.\ 
infall models, there is some scatter in this correlation).  

Now we use this model grid to investigate trends in age and metallicity
with local and global structural parameters in sections 
\ref{c4evo} through \ref{c4sfl}, where we vary the galaxy evolution 
and SFL prescriptions.   A summary of the models presented
in the following sections is given in Table \ref{tab:models}.
The age, stellar metallicity and gas metallicity gradients
are also given for these models in Table \ref{tab:grads}: note
that there may be some slight mismatch in the properties of
the observed and model galaxies (e.g.\ in panel a of Fig.\
\ref{fig:c4phys} the model galaxies are regularly distributed
in a rectangle, whereas the observed galaxies are clustered primarily
towards the centre with a fairly significant contingent of galaxies with
more extreme properties) which may affect the comparison of the 
average model and observed gradients slightly.

\begin{table*}
\begin{minipage}{14cm}
\begin{center}
  \caption[Model descriptions]
	{Models presented in this paper }
  \label{tab:models}
  \begin{tabular}{l|ccccc|l}
  \hline
   Model  & SFL & $n$ & $k$ & Threshold & Yield & Chemical Evolution  \\
  \hline
   Fiducial & Schmidt & 1.6 & 0.012 & - & 0.02 & closed box \\
   I       & Schmidt & 1.8 & 0.012 & - & 0.02 & galaxy mass and radius dependent infall \\
   O        & Schmidt & 1.6 & 0.012 & - & 0.03 & more outflow from 
	smaller galaxies \\
   E        & Schmidt & 1.7 & 0.012 & - & 0.03 & mass dependent formation epoch \\
   T       & Schmidt & 1.8 & 0.012 & Y & 0.02 & infall case I  \\
   D        & Ke98 & 1.0 & 0.030 & - & 0.02 & closed box  \\
  \hline
  \end{tabular} 
\end{center}
\end{minipage} 
\end{table*}

\subsection{How we compare the models and the data} \label{c4comp}

In the subsequent sections, we compare the colour-based ages
and metallicities of a sample of face-on spiral 
galaxies with colour-based ages and metallicities from 
a suite of simple galaxy evolution models.  However, earlier we
stated that the colour-based ages and metallicities were
only robust in a {\it relative} sense.  This leaves open an 
important issue: how should we approach the comparison
of the model ages and metallicities and the data?

To understand how we should approach these model 
comparisons, we need to go back to understanding the nature of the
data.  In essence, the trends in the data of e.g.\ Fig.\ \ref{fig:c4sc}
are describing trends in colour with galaxy properties.  Therefore, 
if the black model points describe well the trends in 
age or metallicity with galaxy properties, then what the model really does is 
adequately describe the trend in optical--near-IR colour with 
galaxy properties.  The model, of course, has one significant
limitation: it interprets the trends in colour solely in terms of
smoothly varying SFHs under the assumption of a constant IMF.  
The data, in contrast, has contributions to the colours from
bursts of star formation, dust, and possibly from variations in the
IMF.  

Low-level bursts of star formation (e.g.\ variations of a factor
of two in SFR over 0.5 Gyr timescales; Rocha-Pinto et al.\ 2000b)
are relatively unimportant: these bursts only contribute modest
extra scatter.  IMF variations between different types of 
galaxy are disfavoured (see e.g.\ Kennicutt 1998; Bell \& de Jong, 
in preparation).  The overall IMF normalisation 
produces subtle effects, as the colours will change 
only very little for plausible changes inm the IMF, but the 
galaxy evolution will be affected through e.g.\ the fraction of
mass locked up in long-lived stars ($1 - R$).  Therefore, IMF
uncertainties are unlikely to affect the qualitative 
behaviour of our models but may require the adoption of e.g.\ 
slightly different SFL parameters such as $n$ or $k$.  Similar
effects are expected for some types of model uncertainty or 
model age differences.

Dust is one major remaining uncertainty.  The amounts and effects 
of dust are still hotly debated \cite{ddp,pelt92,hu94,tully98,k98}, 
so being quantitative
about the effects of dust is difficult.  The zeroth
order expectation is that dust is unlikely to be 
important for e.g.\ low surface brightness or luminosity galaxies, 
and is much more important for higher luminosity or surface 
brightness galaxies (e.g.\ Tully et al.\ 1998).
In this scenario, young and metal poor galaxies would be little 
affected by dust; however, the older and more metal-rich galaxies would
be likely to appear even older and more metal-rich, if dust
was properly accounted for.

For these reasons, it is not fair to take the details of the 
model comparisons with the data too seriously.  However, the 
relative trends shown by the comparison should be reasonably
robust and it is certainly fair to compare the performance of 
different models.  Also, along the same vein, we compare
the models on a qualitative, visual level because
no simple statistical approach (e.g.\ a two dimensional 
Kolmogorov-Smirnov test or least-squares line fitting) can 
take into account both the modelling uncertainties that
we have discussed above and the role of selection effects
in limiting the area of parameter space that is observed.  We show
all of the relevant plots for each model so that the reader
can make their own comparisons and assessments of the different models.

\section{Closed box model} \label{c4evo}

Our fiducial model is a closed box model (i.e.\ $E = F = 0$ in
equations 2 through 4) with a Schmidt 
SFL, with efficiency $k = 0.012$
M$_{\sun}$\,pc$^{-2}$\,Gyr$^{-1}$ at a surface density of 
1 M$_{\sun}$\,pc$^{-2}$ and $n = 1.6$ (see Table \ref{tab:models}; 
note that modest increases in $n$ are 
possible if $k$ decreases, and vice versa).
The primary prediction of the closed box Schmidt SFL model
is that {\it the SFH and metallicity of a given area in 
any galaxy depend only on the initial local gas surface density}.
Note that the star formation 
and chemical enrichment history of a galaxy are {\it independent} 
of the initial density of gas if $n = 1$, i.e.\ if the star formation rate is 
directly proportional to the gas density.   

When we evaluate the effects of different galaxy evolution 
prescriptions, we will evaluate them with respect to this 
model.  As such, it is worth spending some time on understanding 
what this model can and cannot reproduce, in terms of the 
trends observed in BdJ.

\begin{figure}
\begin{center}
\psfig{figure=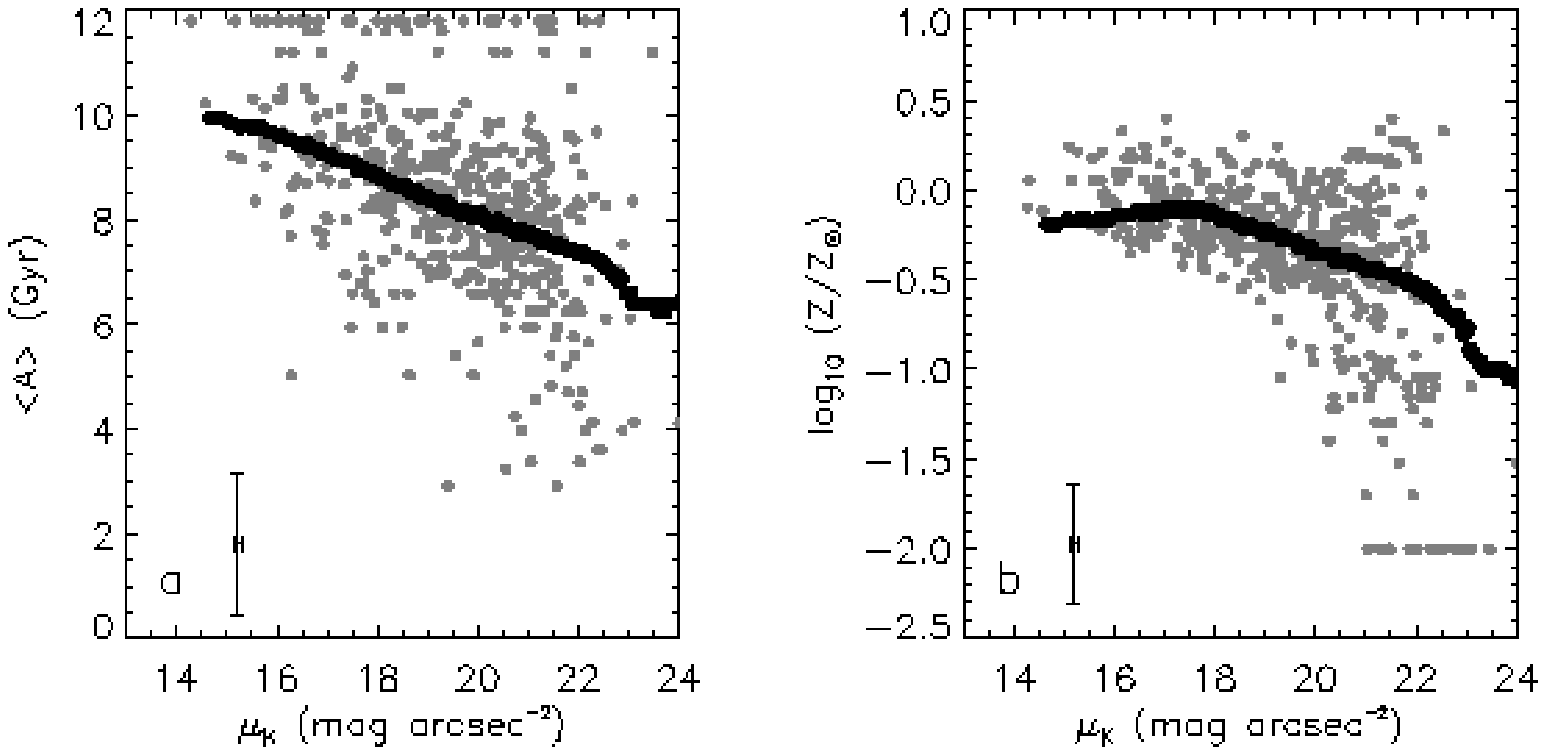,width=8cm}
\psfig{figure=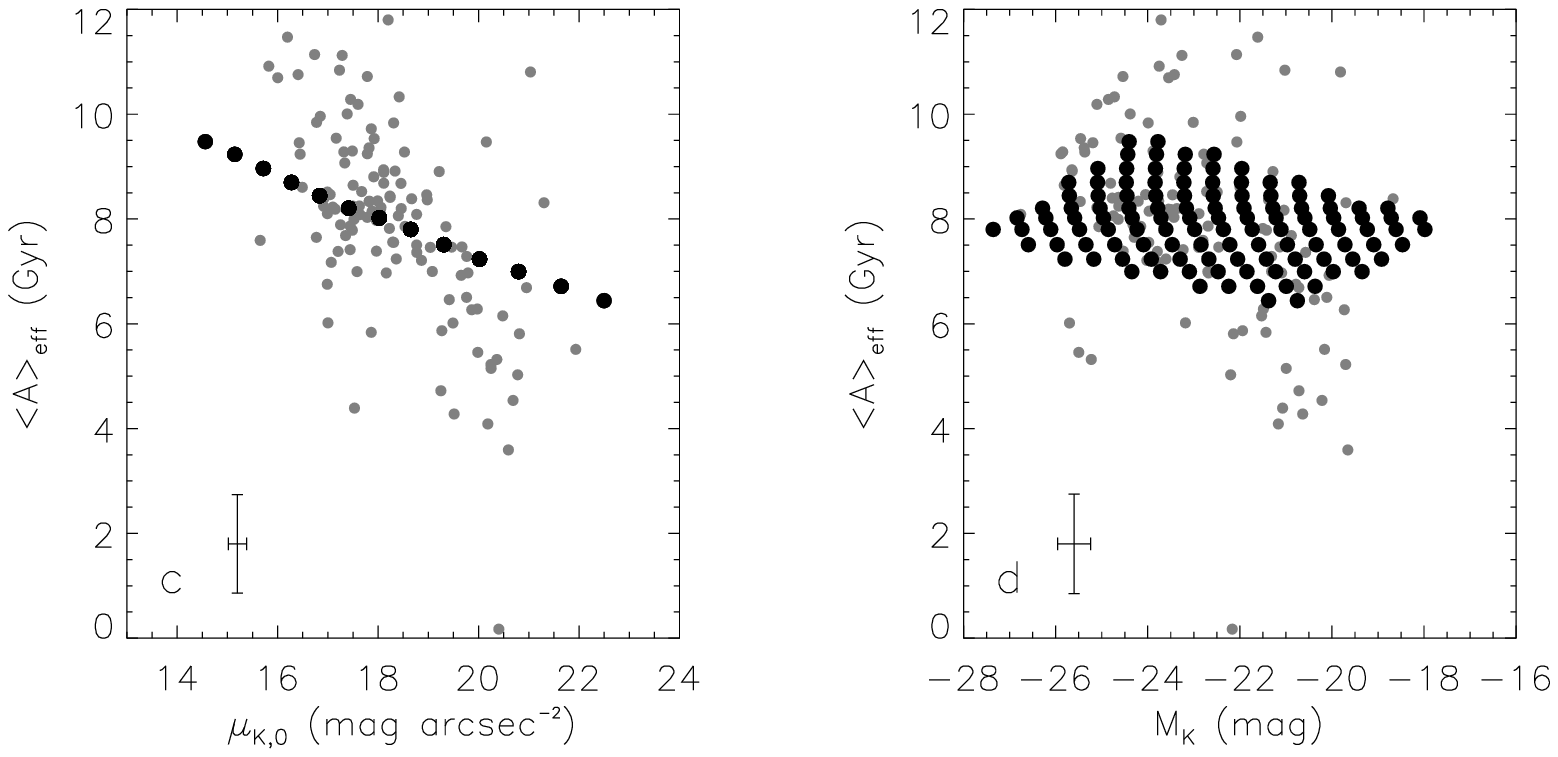,width=8cm}
\psfig{figure=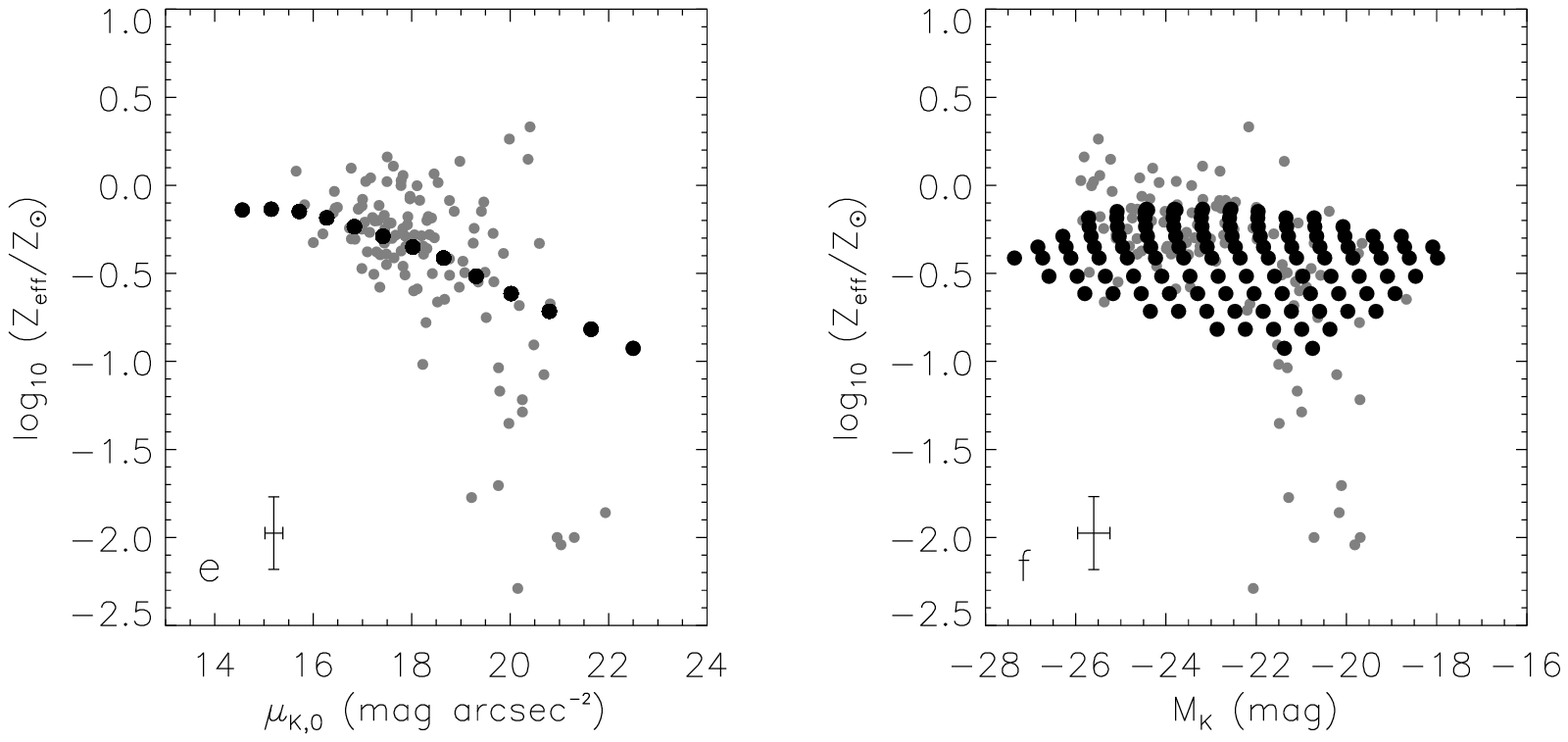,width=8cm}
\psfig{figure=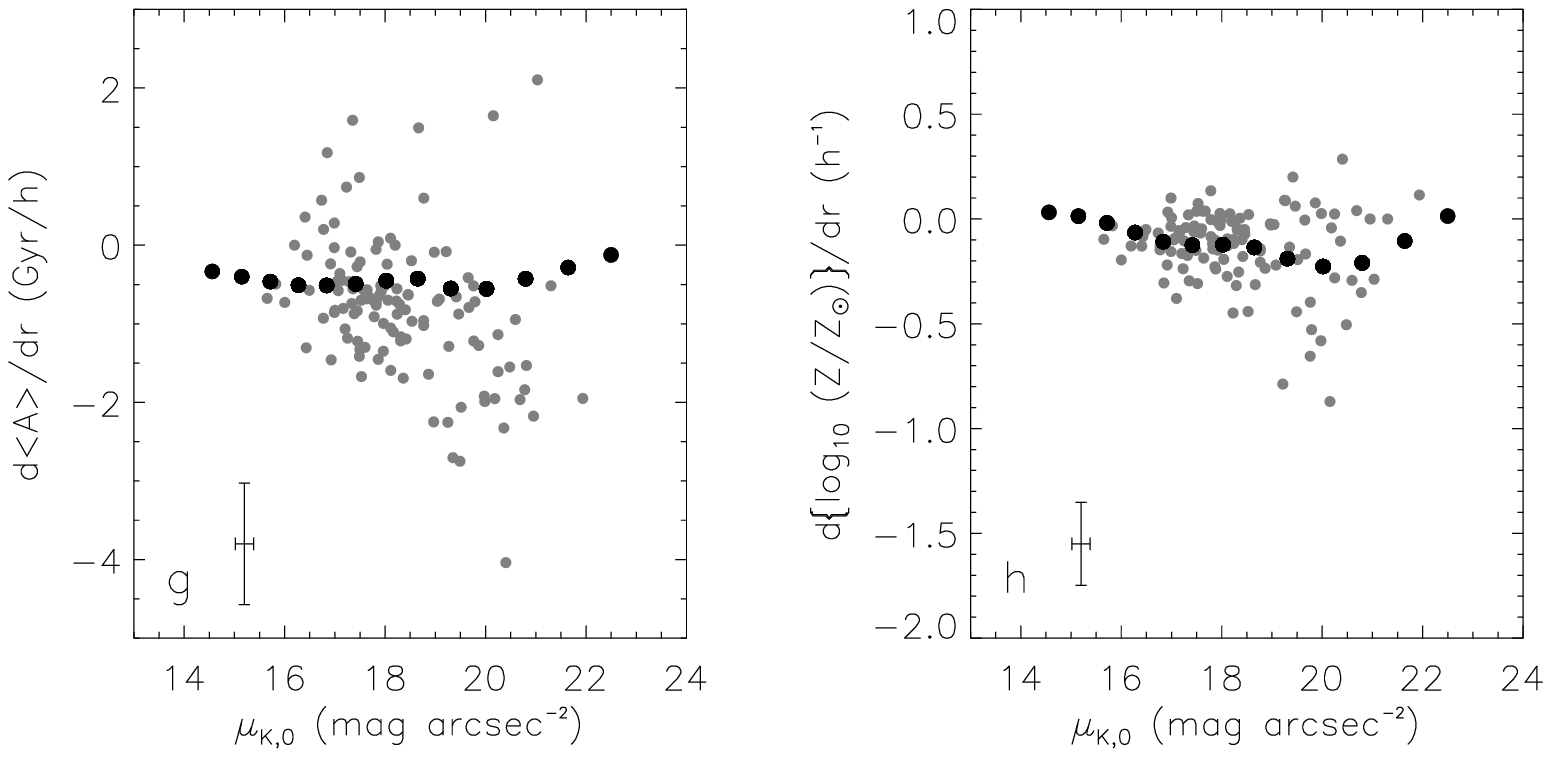,width=8cm}
\end{center}
\vspace{-0.4cm} \caption{ {\bf Fiducial Model: }
Panels a and b show the correlation between local $K$ surface brightness and 
the local age (panel a) and metallicity (panel b) of the data from 
BdJ (grey circles; average error bars are in the bottom left corner) 
and the fiducial model (black 
circles).
Panels c through f show correlations between the ages at the disc half-light
radius and the $K$ band disc central surface
surface brightness $\mu_{K,0}$ (panel c) and the $K$ band 
absolute magnitude $M_K$ (panel d), and
between the metallicity at the disc half-light
radius and the $K$ band disc central surface
surface brightness (panel e) and the $K$ band 
absolute magnitude (panel f).
In panels g and h we show correlations between the age (panel g) 
and metallicity (panel h)
gradients per $K$ band disc scale length and 
the $K$ band central surface brightness.  
In panels a and b, a significant number of the grey
data points from BdJ have average ages of 
11.9 Gyr or metallicities of $\log_{10}(Z/Z_{\sun}) = -2$: these
data points fall outside the model grids. 
The model grid 
becomes quite uncertain at very low metallicities: many of the
grey data points in panels b, e and f at metallicities lower than 
$\log_{10}(Z_{eff}/Z_{\sun}) = -1$ may not have metallicities
as low as those plotted on this diagram.
Some of the older, very low surface brightness galaxies 
may have significant age errors in panels c and d (Bell et al.\ 2000). }
\label{fig:c4sc}
\end{figure}

In Fig.\ \ref{fig:c4phys} we show the fiducial model
selection box and $K$ band central surface brightness--gas
fraction correlation.  As discussed in the last section, 
these two plots are used to select the galaxy models (panel a)
and help to constrain the SFL parameters (panel b).  
In Fig.\ \ref{fig:c4sc} we show the
local age (panel a) and local metallicity (panel b) against the local $K$ band
surface brightness.  These local ages, metallicities and 
surface brightnesses are for individual one scale length wide annuli in 
the sample and model galaxies: in the limit of 
a SFL that depends on gas density only (such as the Schmidt law),
these correlations should be unique, well-defined lines with no 
intrinsic width (e.g.\ the black model points).  

In panels a and b of Fig.\ \ref{fig:c4sc} we see that the fiducial model
does a good job of reproducing the observed trends in age and metallicity
with local $K$ band surface brightness.  
Although age and metallicity increase monotonically with surface brightness
in the model, the ages (or metallicities) 
derived from the colours show more irregular trends with surface 
brightness.  This is primarily due to
degeneracies and irregularities
in the colour-colour grid for stellar populations with near-constant
star formation rates (i.e.\ average ages $\sim 6$ Gyr), which cause
irregularities in the age--surface brightness and metallicity--surface
brightness plots.  

However, the observational data have significant
scatter: while this is partly due to observational errors, there is a 
component of the scatter which is real intrinsic scatter in 
annulus colour at a given surface brightness.   This indicates 
at least departures from a smooth SFH 
(e.g.\ Rocha-Pinto et al.\ 2000b), and potentially
indicates dependence on some factor other than the local 
$K$ band surface brightness (BdJ).  Thus, while the model gives
a good match to the overall trend, additional factors need to be introduced
to account for the scatter between annuli with the same $K$ band
surface brightnesses.

Recall that we carried out linear fits to trends in age and 
metallicity with radius within a galaxy (yielding an intercept
at the half-light radius and a gradient per $K$ band disc scale
length).
In panels c and d of Fig.\ \ref{fig:c4sc}
we explore the correlations between the age 
intercept at the half-light radius and the $K$ band central 
surface brightness (panel c) and absolute magnitude (panel d).  
We see that the fiducial model produces
a correlation between the age at the half-light radius 
and the $K$ band central 
surface brightness (panel c: also between age and gas fraction; not shown).
Note however that the slope of the model correlation is too shallow:
real galaxies
show a steeper correlation between age at the half-light radius and 
$K$ band central surface brightness than the model galaxies.

The selection limits imposed on the model galaxies 
produce a small residual correlation between age at the 
half-light radius and $K$ band absolute magnitude (panel d): 
this correlation is 
fictitious (age does not depend on mass in this model) and
is the result of the magnitude--surface brightness correlation shown in
panel a of Fig.\ \ref{fig:c4phys}.  The correlation imposed
by the magnitude--surface brightness correlation on 
panel d of Fig.\ \ref{fig:c4sc} is too shallow compared to the data,
however.

In panels e and f of Fig.\ \ref{fig:c4sc} 
we explore correlations between the
metallicity intercept at the disc half-light radius and the 
$K$ band central surface brightness (panel e) and absolute magnitude
(panel f).  
Note that the model grid becomes quite
uncertain at low metallicities:  many of the grey data points
at metallicities lower than $\log_{10}(Z_{eff}/Z_{\sun}) = -1$ may
not have metallicities as low as those plotted (BdJ).
The metallicity--surface brightness correlation (panel e: 
and metallicity--gas fraction correlation; not shown)
is well-described by 
the fiducual model.  We can also see from panel f of 
Fig.\ \ref{fig:c4sc}
that there is a small residual correlation between metallicity
and magnitude imposed by the selection limits in panel a of 
Fig.\ \ref{fig:c4phys}, 
but that the slope of the correlation is too small:  this indicates
that metallicity has both a surface brightness and magnitude dependence
(see e.g.\ BdJ; Skillman, Kennicutt \& Hodge 1989; Garnett et al.\ 1997).

In panels g and h of Fig.\ \ref{fig:c4sc} and Table \ref{tab:grads}, 
we explore the age (panel g) and metallicity
(panel h) gradients per $K$ band disc scale length.  
Note that these gradients are considerably noisier than 
the intercepts that were discussed above: in particular, 
individual metallicity gradients are often only marginally detected (see
BdJ for an indication of the uncertainties, and versions of
these data plots with error bars).  

Both the amplitude of and trends in age gradient with 
$K$ band central surface brightness are poorly reproduced by the fiducial
model (panel g of Fig.\
\ref{fig:c4sc}): there is little if any correlation between 
age gradient and surface brightness in the model, and the model 
underpredicts the average 
age gradient in the data by around a factor of two
(see Table \ref{tab:grads}).  This problem may be linked
with the inability of the model to reproduce the steepness of the 
global age--$K$ band central surface brightness correlation in 
panel c of Fig.\ \ref{fig:c4sc}, as the slope of the model correlation 
is too shallow, indicating that the rate of change
of age with surface brightness (i.e.\ the age gradient) is 
underpredicted by the model.  In contrast, the fiducial 
model reproduces both the average
metallicity gradient and the trends in metallicity gradient with $K$ band
central surface brightness (albeit with no scatter in the model metallicity
gradients; 
Table \ref{tab:grads} and panel h of Fig.\ \ref{fig:c4sc}).  

To summarise, the fiducial model does a reasonable job of describing
the correlations between the ages and metallicities of galaxies and 
their physical parameters.  The main shortcomings of the fiducial model
are a lack of magnitude dependence in both the age and metallicity 
(panels d and f of Fig.\ \ref{fig:c4sc})
and the underprediction of the rate of change of age with $K$ band 
surface brightness (Table \ref{tab:grads} and panels c and g of Fig.\ 
\ref{fig:c4sc}).  
An additional shortcoming which we address more
explicitly later in section \ref{c4disc} 
is that closed box models predict too many 
low metallicity stars in both the solar neighbourhood and in 
external galaxies (the G dwarf problem; e.g.\  
Rocha-Pinto \& Maciel 1996; Worthey, Dorman \& Jones 1996; Pagel 1998).
In the next section, we see how 
modifying the fiducial model by introducing infall, outflow 
or systematic trends in galaxy formation epoch
can improve the match between the models and the observed correlations.

\section{Galaxy evolution prescriptions} \label{c4inf}

Can any of these shortcomings of the fiducial model be
alleviated by invoking infall, outflow or variations in formation 
epoch between galaxies?  
Infall is a natural part of galaxy formation: 
at some level the disc of a spiral galaxy must be built up 
by infall, and a plausible explanation for the local G dwarf problem
is an extended infall history at the solar cylinder 
(e.g.\ Tinsley \& Larson 1978; Lacey \& Fall 1983; 
Pagel 1998; Boissier \& Prantzos 1999).
Outflow is somewhat more speculative: successful heirarchical models
of galaxy formation rely on negative feedback to supress star formation 
in low-mass systems at early times.  Without feedback, these models
generically produce too many faint galaxies to be compatible with
the observed luminosity function (e.g.\ 
Kauffmann \& Charlot 1998; Somerville \& Primack 1999; Cole et al.\ 2000).
The most likely source of feedback is the energy released by Type
II supernovae.  The feedback may take several forms: e.g.\ it may result
in the preferential ejection of high metallicity gas or it may
result in a `super-wind' which completely removes the gas content
of the galaxy (e.g.\ Dekel \& Silk 1986; Arimoto \& Yoshii 1987; 
MacLow \& Ferrera 1999; Martin 1999).  
Significant differences in formation epoch 
(i.e.\ significant differences in the time
at which a galaxy's gas supply becomes available for star formation) 
between galaxies of e.g.\ different masses or halo spin parameters are quite
possible.  Semi-analytic, linear collapse and gas-dynamical cosmological
simulations all suggest systematic trends in galaxy formation epoch with halo
properties (e.g.\ Cole et al.\ 2000; Mo, Mao \& White
1998; Steinmetz \& M\"{u}ller 1995).
We explore the
effects of infall, outflow and variations in formation epoch below in the
next three subsections.

\subsection{Infall} \label{c4infall}


\begin{figure}
\begin{center}
\psfig{figure=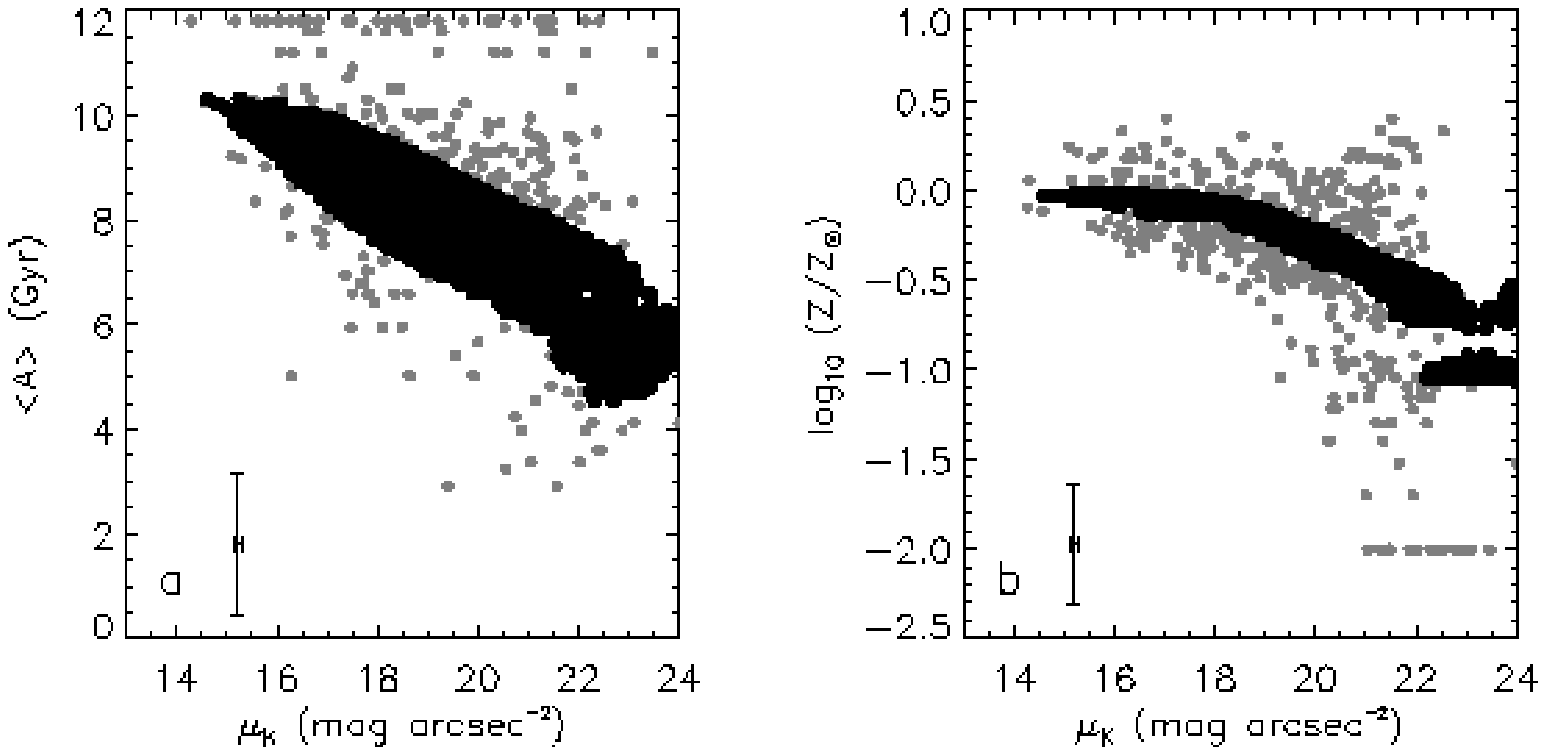,width=8cm}
\psfig{figure=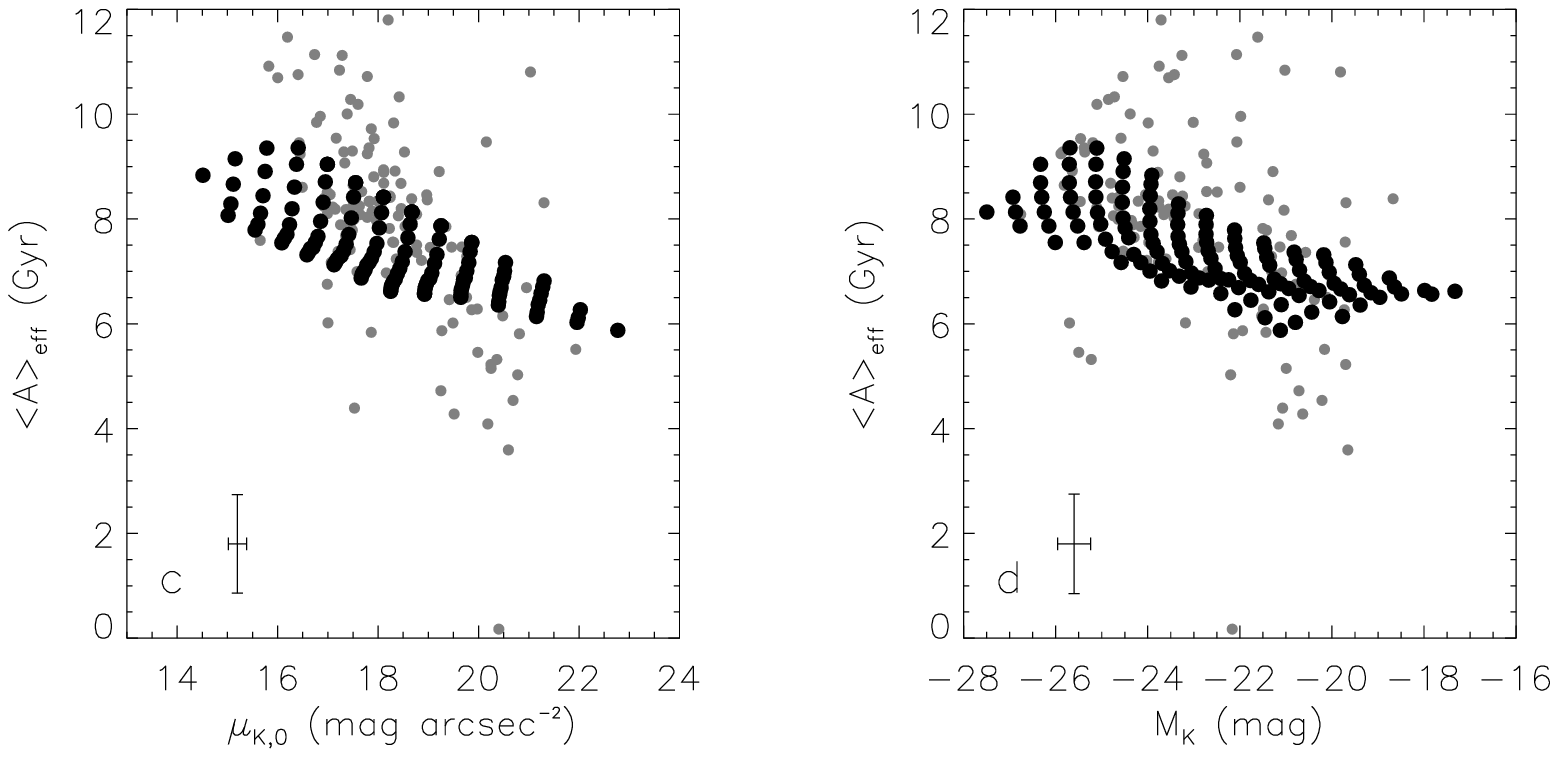,width=8cm}
\psfig{figure=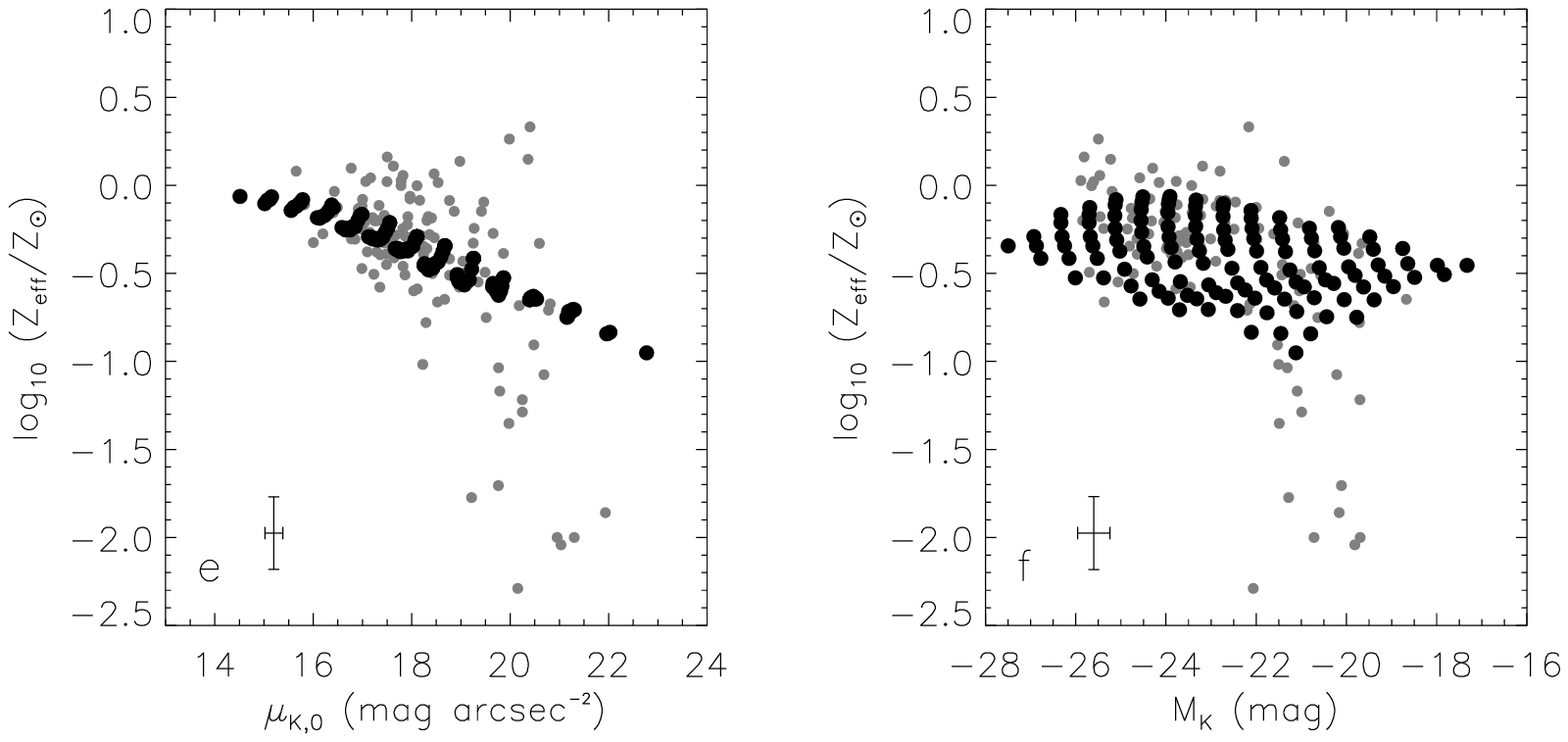,width=8cm}
\psfig{figure=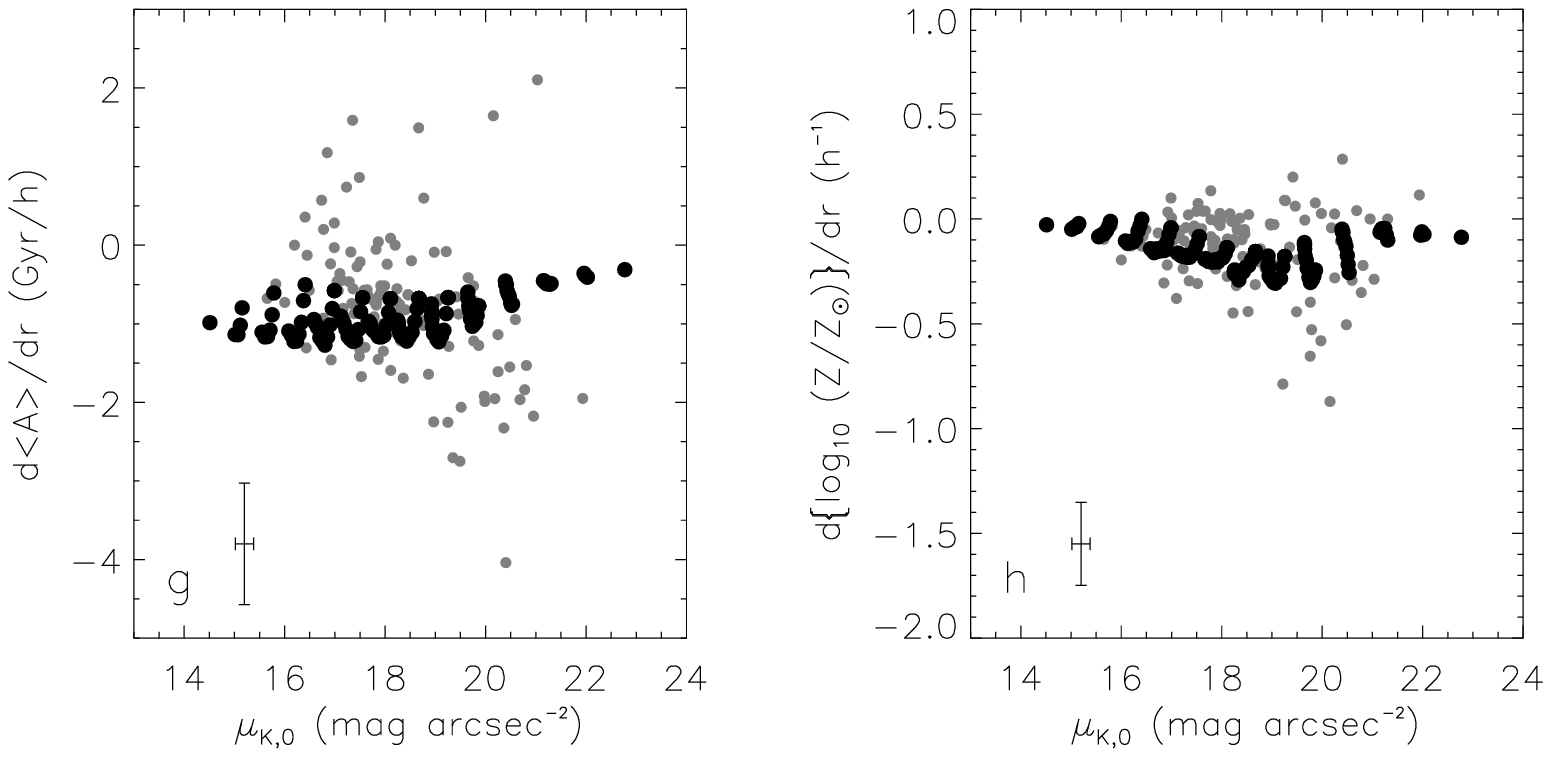,width=8cm}
\end{center}
\caption{{\bf Infall Model I: }
Panels and symbols are as in Fig.\ \protect\ref{fig:c4sc}}
\label{fig:c4inf}
\end{figure}

In order to test the effects of infall, we adopt a simple parameterisation.
We assume that the infall timescale varies with both galaxy 
mass and radius within a galaxy: a radial dependence, 
and to a certain extent, a mass dependence
in infall timescale are expected theoretically 
(e.g.\ Eggen, Lynden-Bell \& Sandage 1962; 
Larson 1976; Contardo et al.\ 1998; Mo, Mao \& White 1998; 
Boissier \& Prantzos 2000; Cole et al.\ 2000).
We have also run cases with only mass-dependent infall timescales,
and even a mass-independent infall timescale;
however, these cases are not discussed further as the effects of each of 
these individual cases are included (and are easily
distinguished) in the infall model we consider.

We assume that the galaxy initially has a surface density of zero 
at all radii:  we then build up the exponential disc of the galaxy from 
a reservoir of metal-free gas with an exponential timescale
$\tau_{\rm infall}$.  The infall timescale depends on both 
mass and radius.  The timescale $\tau_{\rm infall}$ 
ranges from $\tau_{\rm infall} = 8$ Gyr
for galaxies with a total mass of $10^{10}$ M$_{\sun}$ to
$\tau_{\rm infall} = 0.5$ Gyr for galaxies with a total 
mass of $\ge 10^{13}$ M$_{\sun}$ according to the formula:
$\tau_{\rm infall} = 8 - 2.5 \log_{10}(M_{galaxy}/10^{10}
{\rm M_{\sun}})$ for galaxies with masses lower
than $M_{galaxy} = 10^{13}$ M$_{\sun}$; for galaxies
more massive than $10^{13}$ M$_{\sun}$ the infall timescale
is fixed at 0.5 Gyr. 
Thus the most massive galaxies build up their
gas mass quickly, and then evolve almost as a closed box, while low-mass 
galaxies have only recently reached their maximum gas mass and star
formation rate.  In this model, the infall starts at the same epoch 
in every galaxy and is always a decaying exponential: we investigate the 
effects of varying the formation epoch in Section 4.3.

The above infall timescales are defined
at the half-light radius $R_e$.  To impose radial 
variation, we reduce the central
infall timescale by a factor of two and the infall timescale 
at $2R_e$ is doubled.
As an example, a $10^{12}$ M$_{\sun}$ galaxy in
this model has infall timescales at $(0,1,2)R_e$ of $(1.5,3,6)$ Gyr.

All the model galaxies 
with infall have a Schmidt SFL with exponent $n = 1.8$ and 
efficiency $k = 0.012$.   Note that these models
have a different Schmidt law exponent from the fiducial model:  
this is because infall delays star formation compared
to the closed box model, therefore 
star formation must be more efficient than in a closed
box to provide a satisfactory match to the age observations.

From Fig.\ \ref{fig:c4inf} 
(comparing it to Fig.\ \ref{fig:c4sc} for the fiducial model) we
can see that infall mainly affects the ages
of the model galaxies.  The reduction 
of the average age depends (of course) on the infall timescale:
the maximum average age of a stellar population with 
an exponential infall timescale $\tau_{\rm infall}$ decreases as the 
infall timescale increases (and is roughly $12 - \tau_{\rm infall}$ Gyr for
$\tau_{\rm infall} \la 4$ Gyr).  Metallicity is affected less: 
the difference between the mean stellar metallicity of a closed box and 
an exponential infall timescale model at a given gas fraction (therefore
roughly surface brightness) is $\la 0.1$ dex for any plausible SFL.
Infall does `narrow' the
metallicity {\it distribution} of a stellar population, where there
are relatively less low and high metallicity stars than the 
closed box model: this in fact was one of the original motivations
for the infall model (e.g.\ Larson 1972; 
Tinsley 1980; Prantzos \& Aubert 1995; Pagel 1998; Prantzos \& Silk 1998).

The mass dependence in the infall timescale introduces
a slope in the magnitude--age intercept relation 
(panel d in Fig.\ \ref{fig:c4inf}).  The slope is too shallow 
to adequately describe the trends in the data; however, this
is closely linked to the underprediction of the slope of the
age intercept--central surface brightness correlation
(panel c in Fig.\ \ref{fig:c4inf}) which
feeds into a shallow slope
for the age intercept--magnitude correlation.  Tuning a better
match with an infall model is difficult: the real problem is achieving
a stronger variation of age with surface brightness which 
is still consistent with the data in panels a and c of Fig.\ \ref{fig:c4sc}.
Perhaps this could be linked to 
a central surface density dependent infall history (e.g.\ through 
a spin parameter dependence in halo formation timescale); however,
adding yet another parameter to the infall model is unappealing 
both in terms of the simplicity of the modelling technique and 
the quality of the data at this stage.

The radial dependence in the infall timescale increases the age 
gradient: this is illustrated in panel g of Fig.\ \ref{fig:c4inf} and
Table \ref{tab:grads}, where the
metallicity gradients (both gas and stellar) are relatively
unchanged from the closed box case, but the age gradient is 
increased by a factor of two.  Both the radial and mass dependence
increases the scatter in the local age--local $K$ band
surface brightness relation, but not towards unacceptable levels.

To summarise: infall affects primarily the age of a region
within a galaxy, but does not significantly affect the mean metallicity
of the stellar population.  Therefore, mass (or radial) variation
of the infall timescale introduces mass dependence (or a radial gradient)
in the age of a galaxy.

\subsection{Outflow} \label{c4outflow}

\begin{figure}
\begin{center}
\psfig{figure=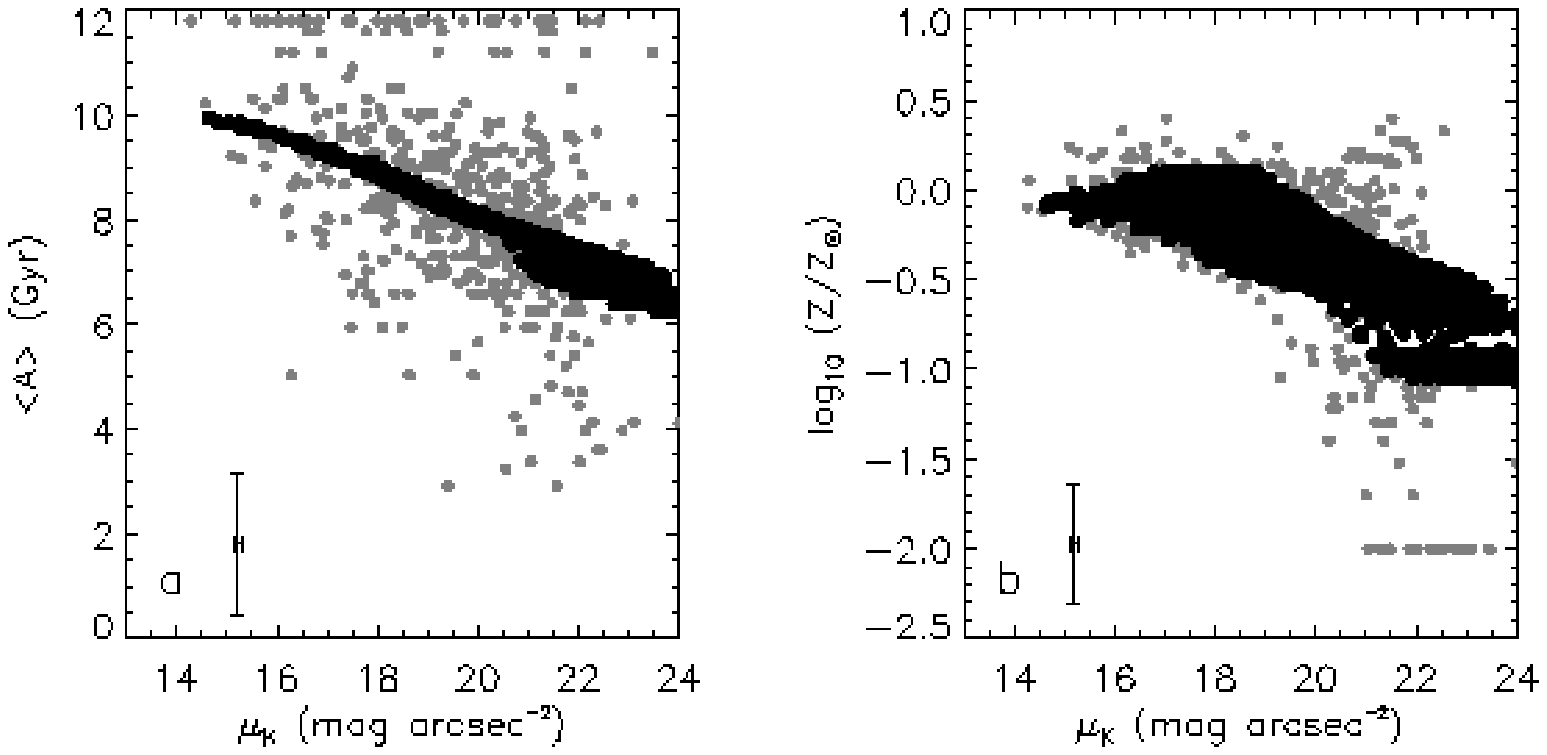,width=8cm}
\psfig{figure=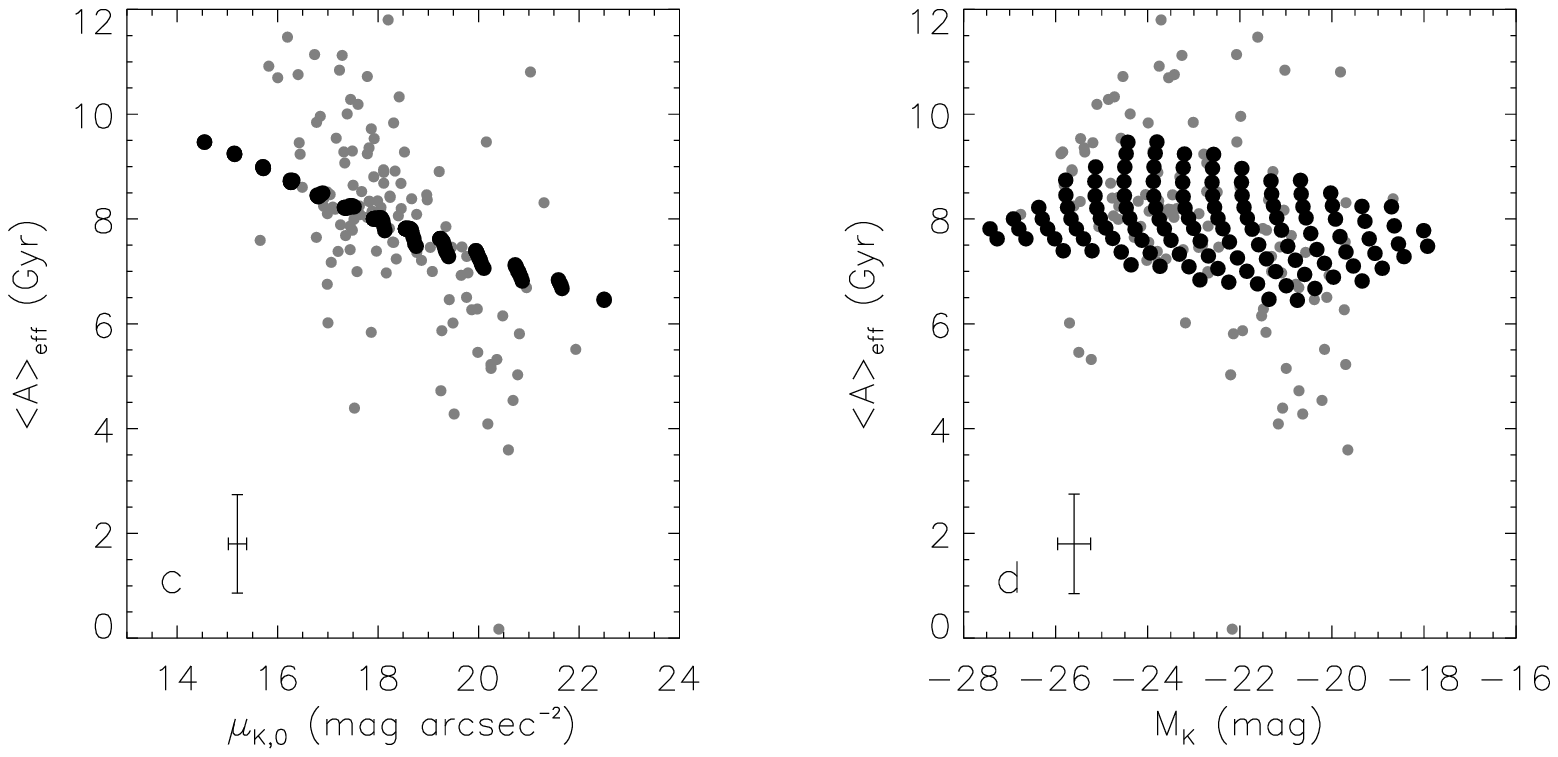,width=8cm}
\psfig{figure=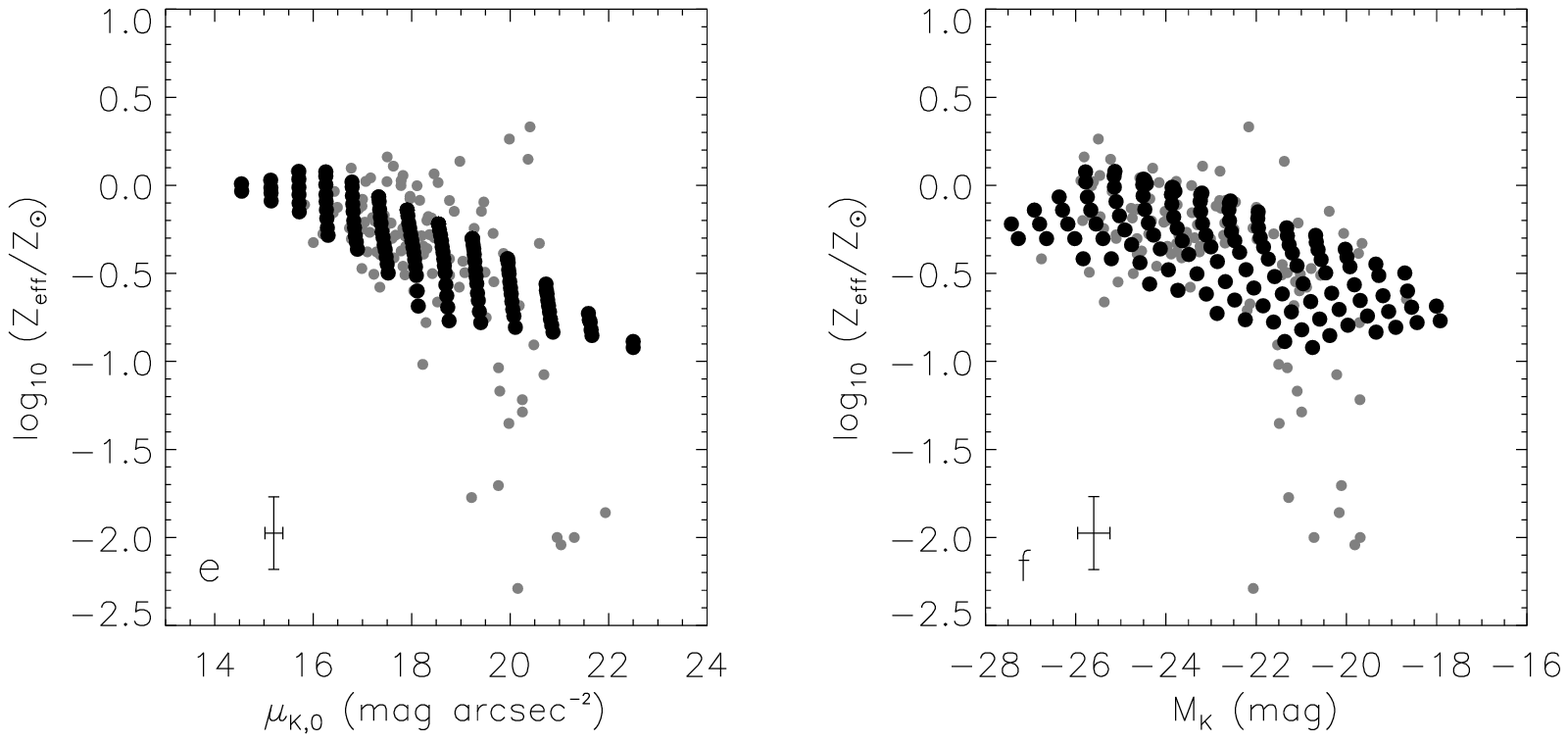,width=8cm}
\psfig{figure=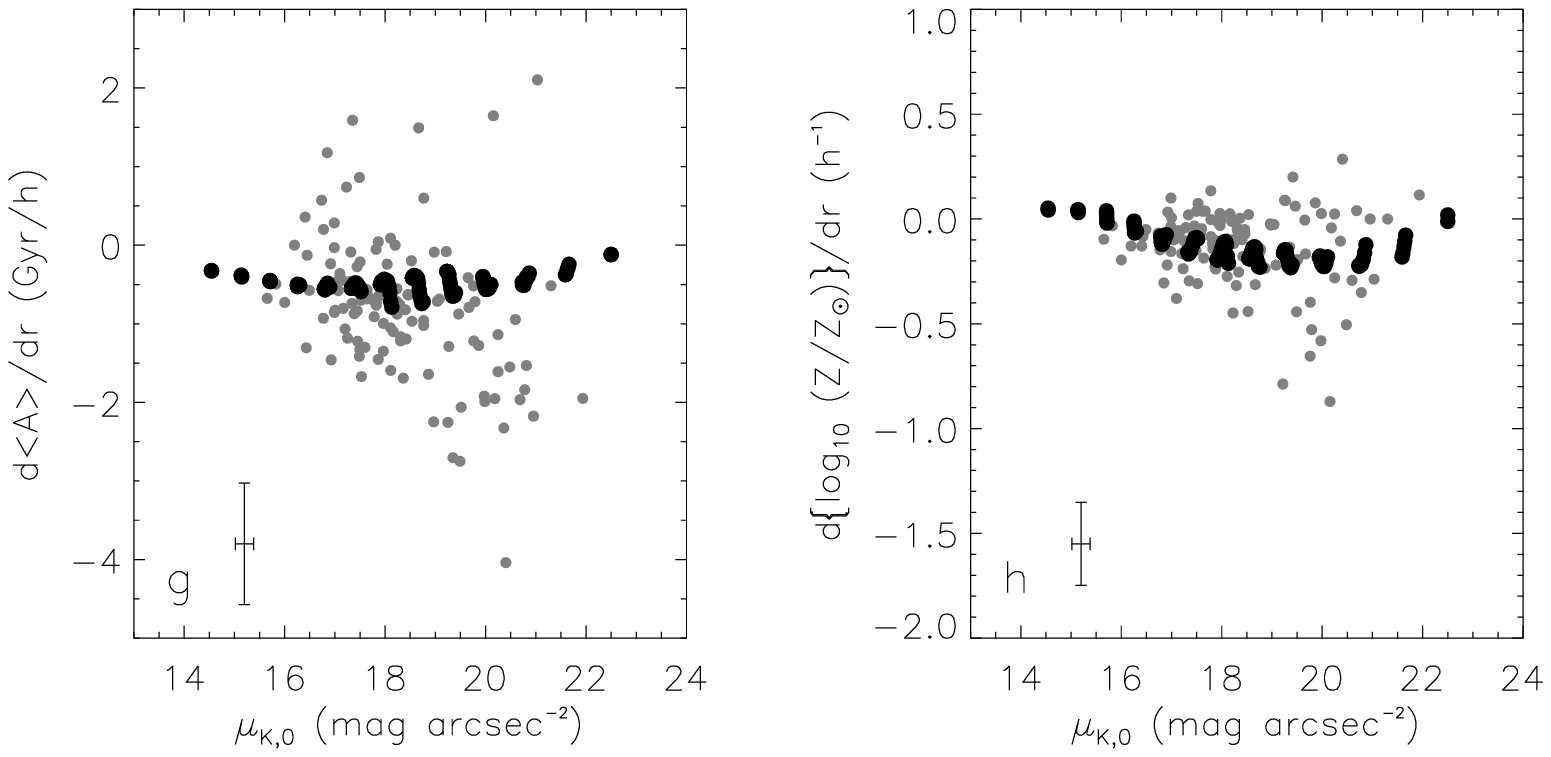,width=8cm}
\end{center}
\caption{{\bf Outflow Model O: }
Panels and symbols are as in Fig.\ \protect\ref{fig:c4sc}}
\label{fig:c4out}
\end{figure}

In order to test the possible importance of outflow, we adopt
a basic parameterisation of its effects.  Outflow caused
by supernovae winds is predicted
to be much more effective for low-mass galaxies (Dekel \& Silk 1986; 
Arimoto \& Yoshii 1987; 
MacLow \& Ferrera 1999; Martin 1999; Cole et al.\ 2000). Furthermore,
for galaxies with total masses in excess of $10^9$ M$_{\sun}$, there
may be little gas mass loss;  however, many of the freshly-synthesized
metals can be lost reasonably easily (MacLow \& Ferrera 1999).
Accordingly, we adopt this simple outflow recipe:
galaxies with masses greater than $10^{13}$ M$_{\sun}$ lose no 
metals in an outflow; galaxies with masses lower than 
$10^{13}$ M$_{\sun}$ lose increasingly more and more metals
with decreasing mass, 
down to a limit of $10^{9}$ M$_{\sun}$, where a galaxy loses
90 per cent of its freshly-synthesized metal content.
{\it We assume negligible gas mass loss in the outflow} in order
to separate the effects of outflow from those discussed in the previous
section: the fraction $f$ of metal mass is given by $f = 
0.9 - 0.225 \log_{10}(M_{galaxy}/10^{9} {\rm M_{\sun}})$ 
below $M_{galaxy} = 10^{13}$ M$_{\sun}$; for 
$M_{galaxy} \ge 10^{13}$ M$_{\sun}$ we assume no metal-enriched
outflow.
This approximation glosses over all the physics behind outflow; 
however, it does allow us to explore the possible effects of outflow
in a simple, well-defined way.  Note that there is no infall 
in this model.

Results from the outflow model (model O) are shown in Fig.\ 
\ref{fig:c4out}.  
Note that we have increased the true yield $p$ 
from solar metallicity for the fiducial model
to 1.5 solar metallicity for the outflow case: 
as outflow involves the loss of metals from 
the model galaxies, a larger yield must be used to allow the models
to reproduce the mean observed metallicities.  
In contrast with the infall case, 
outflow leaves the colour-based age of a stellar population relatively
unaffected.  This is to be expected since the SFL only depends on 
gas density and since
there is no gas {\it mass} loss in the outflow, the SFHs of
the outflow model are the same as those from the closed box fiducial model.
However, outflow profoundly affects the 
metallicity of a galaxy, causing a great deal of scatter
in the metallicity--local $K$ band surface brightness diagram.
The effects of outflow are visible in panels e and f of Fig.\
\ref{fig:c4out}, where we plot the metallicity at the disc
half-light radius against the galaxy parameters.  Comparison of panels
e and f of 
Fig.\ \ref{fig:c4out} for the outflow model with 
Fig.\ \ref{fig:c4sc} for the fiducial (closed box) model 
clearly shows that outflow of this type produces
a strong mass dependence in the galaxy metallicity; furthermore, 
a simple model of this type reproduces the metallicity scatter fairly 
accurately.

To summarise: outflow affects the metallicity of a galaxy.
If outflow is mass dependent, it can imprint a metallicity-mass correlation
without significantly affecting the age of the model galaxies.

\subsection{Variation in Formation Epoch} \label{c4agediff}

In the previous three models we have assumed that all galaxies start
forming stars at a common epoch.
Heirarchical models of galaxy formation predict that
more massive galaxies, despite being {\it assembled} later (from 
a number of smaller subunits), have the
bulk of their stellar population forming earlier than less massive
galaxies (e.g.\ Somerville \& Primack 1999; Cole et al.\ 2000).
Furthermore, stripped-down linear collapse models (e.g.\
Mo, Mao \& White 1998; Dalcanton et al.\ 1997) as well as the
more comprehensive gas-dynamical simulations (e.g.\ 
Steinmetz \& M\"{u}ller 1995; Contardo et al.\ 1998) 
predict that disc formation must happen relatively late
to produce disc galaxies that have anywhere near enough angular
momentum to match the observations (although note that 
gas dynamical cosmological simulations produce discs that
are typically too small to match the observations, e.g.\
Navarro \& Steinmetz 1997; 2000). 
However, we have assumed in the previous sections
that all galaxies are formed at a look-back time of 12 Gyr.  

To investigate the impact of variations in the formation epoch
\footnote{We reserve the term `formation epoch' to
indicate the look-back time at which stars start to from. We use
the term `colour age' (or simply `age') to mean the average
age of the stellar population as recovered from our colour inversion
algorithm.} 
on spiral galaxy colours, 
we allow the formation epoch of a galaxy to depend on its halo 
mass, where galaxies with masses greater than $10^{13}$ M$_{\sun}$
have form at a look-back time of 12 Gyr, 
and galaxies with masses lower than 
$10^{13}$ M$_{\sun}$ have formation epochs, $E$, 
smoothly decreasing from 12 Gyr at
$M_{galaxy} = 10^{13}$ M$_{\sun}$ to 4 Gyr at $10^9$ M$_{\sun}$ according to 
the formula $E {\rm (Gyr)} = 4 + 2\log_{10}(M_{galaxy}/10^9$ M$_{\sun})$.
This approximation, in the spirit of this semi-empirical modelling, glosses
over the important physics behind these variations in formation 
epoch in an attempt
to gain insight into their most important observable consequences.  
We always assume a galaxy formation epoch of 12 Gyr when 
interpreting the broad band colours in the age and metallicity
fitting routine.  This assumption is necessary: it was 
the assumption used to derive the observational ages and 
metallicities, and in the fitting routine we have no {\it a priori} 
knowledge of the formation epoch of galaxies.  An additional benefit of
this modelling is to demonstrate how much effect different galaxy formation
epochs would have on the observational ages and metallicities that we derive
using this fitting routine.

The variable formation epoch galaxies are assumed to be closed
box systems, with a yield of 0.03 (1.5 times solar) and Schmidt
SFL exponent of $n = 1.7$ and efficiency at 1 M$_{\sun}\,{\rm pc}^{-2}$
of $k = 0.012$ M$_{\sun}\,{\rm pc}^{-2}\,{\rm Gyr}^{-1}$.
The Schmidt SFL exponent was raised to 1.7 and the yield 
was increased to 0.03 to allow the rather younger galaxies
reach gas fractions and metallicities at a given surface brightness
which are in rough
agreement with the observations.  If the same set of 
Schmidt SFL model parameters were used as for the fiducial
model the conclusions would be unchanged but the match to 
the observations would apprear slightly poorer.
No infall or outflow is included in this set of models.

One important caveat: we parameterise the formation epoch model
by simply starting the evolution of the galactic disc at a 
particular look-back time. In practice, it might be more suitable
to think of this model
as allowing large amounts of very late infall. For example, if a small
fraction of stars were formed at early times but the bulk 
of stars were formed only a few Gyr ago or less, we still
parameterise that galaxy by a single, relatively recent
formation epoch.  The basic point is that the formation epoch 
really parameterises
when the bulk of the gas supply became available for star formation: how
that gas supply became available is relatively unimportant from this
perspective.

\begin{figure}
\begin{center}
\psfig{figure=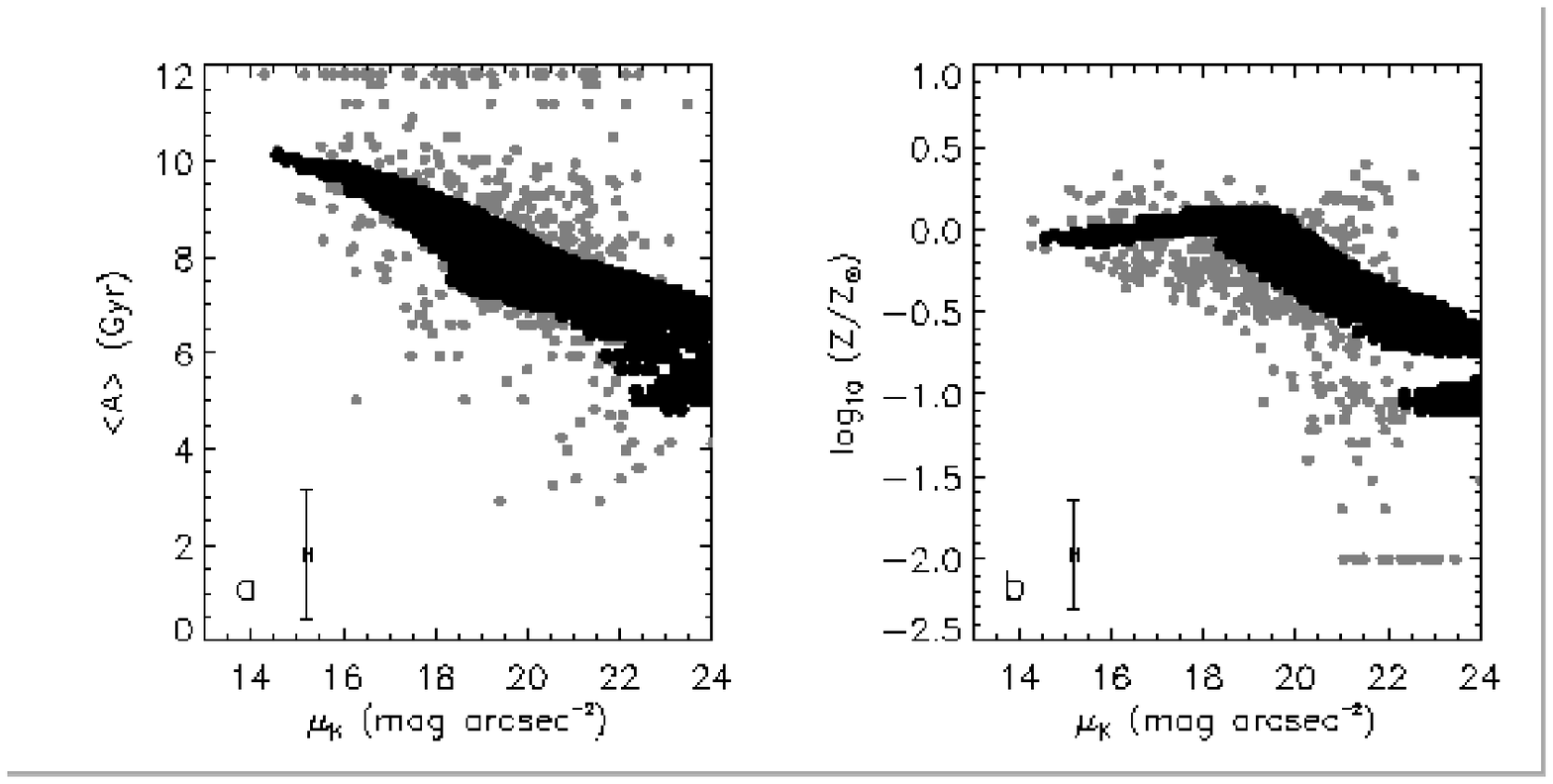,width=8cm}
\psfig{figure=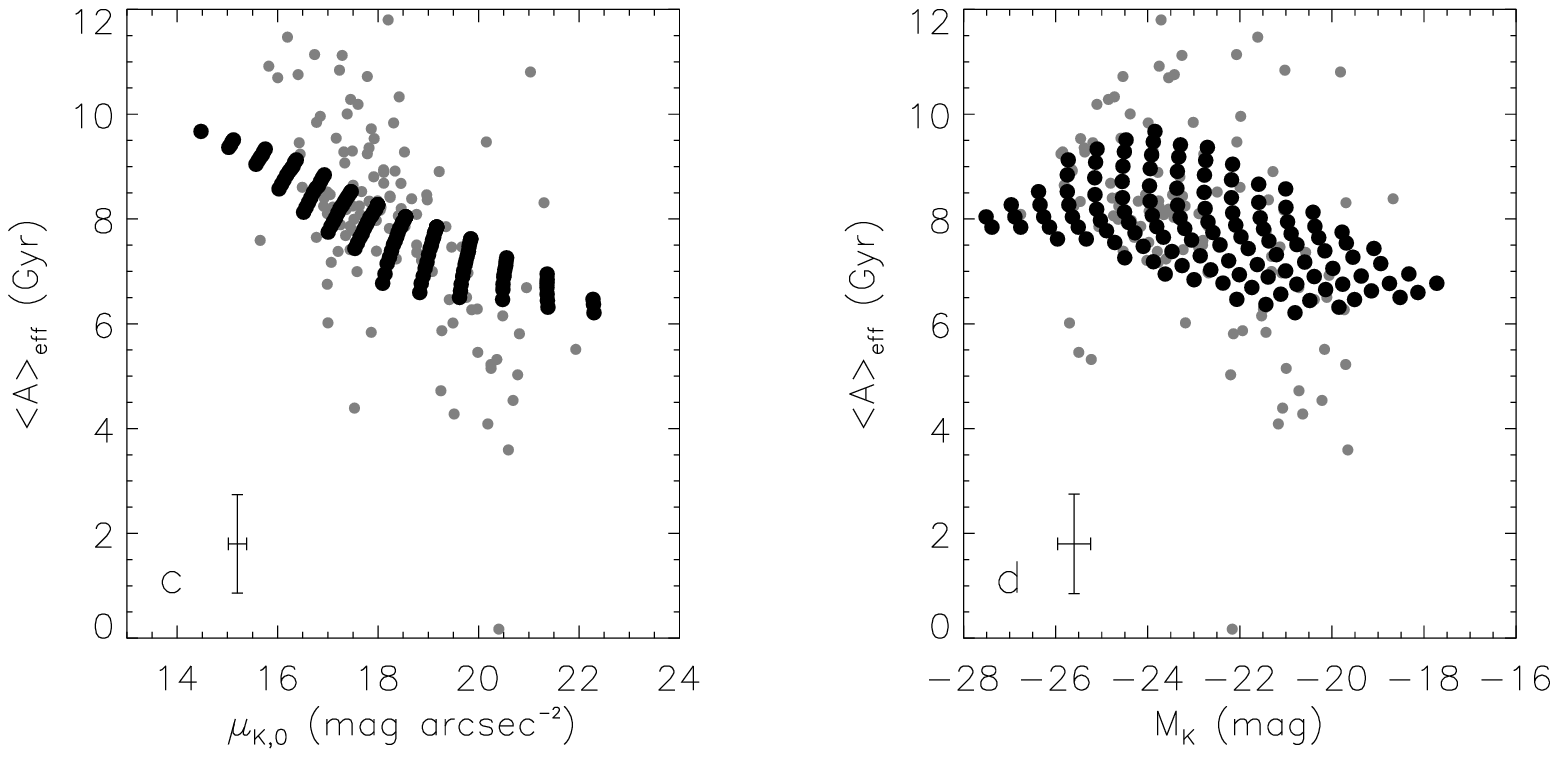,width=8cm}
\psfig{figure=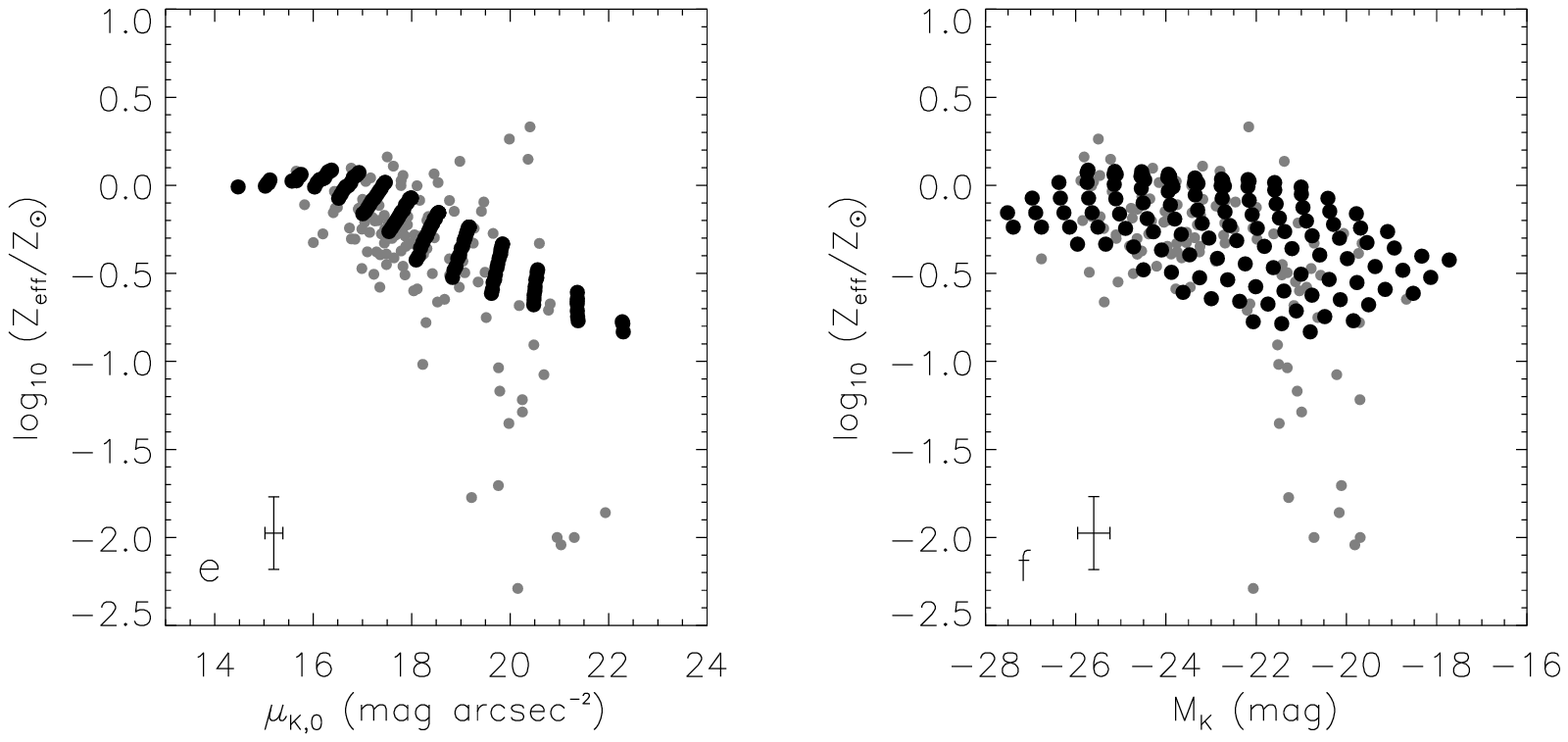,width=8cm}
\psfig{figure=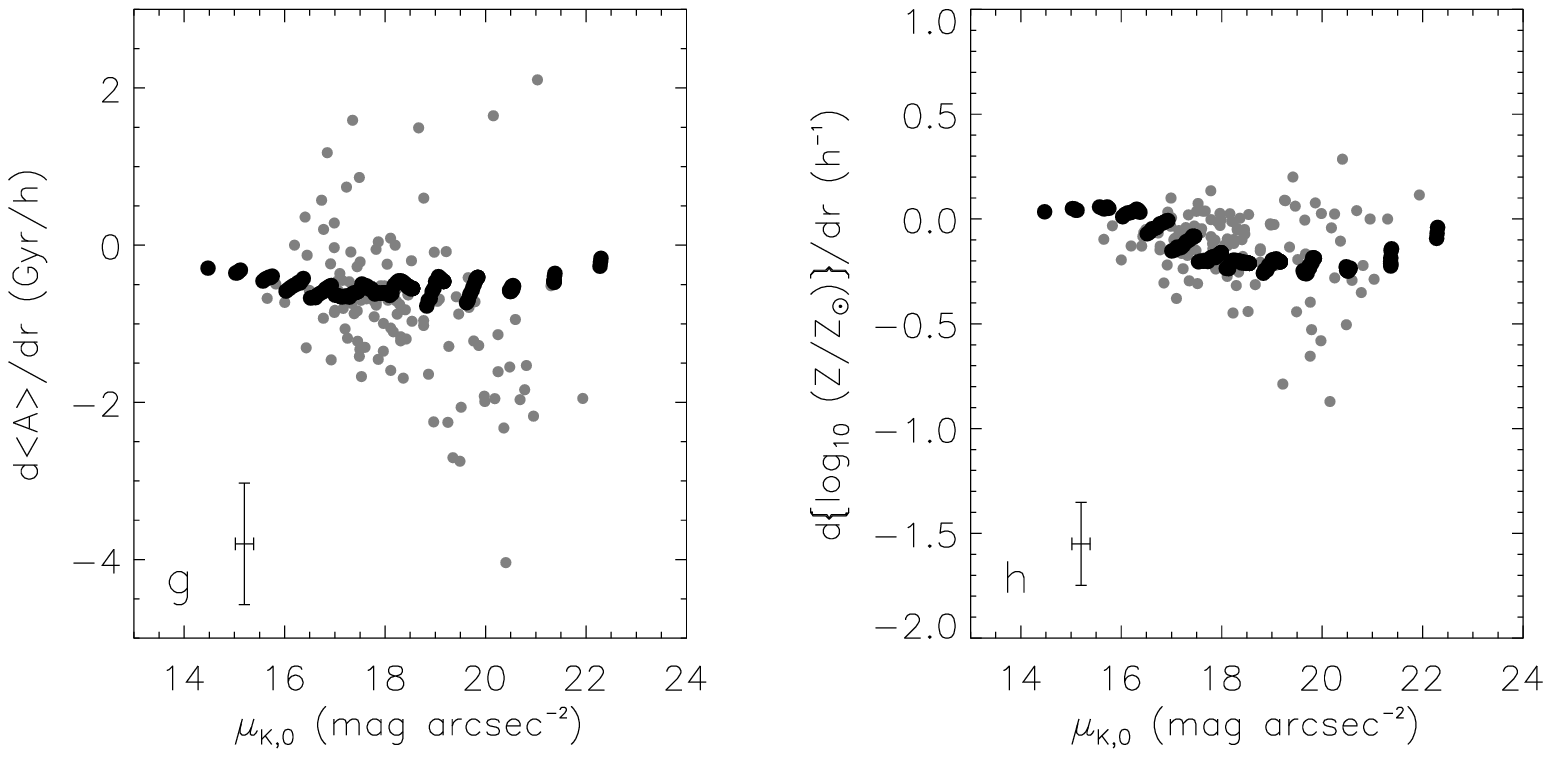,width=8cm}
\end{center}
\caption{{\bf Formation Epoch Model E: }
Panels and symbols are as in Fig.\ \protect\ref{fig:c4sc}}
\label{fig:c4age}
\end{figure}

We show the effects of the mass-dependent formation epoch in Fig.\
\ref{fig:c4age}.  
One important first result is that because our colour-based age estimator
is essentially a birthrate parameter estimator (estimating 
the proportion of young to old stars), it is possible to have large
apparent galaxy ages for galaxies which are, in fact, relatively
young.  As an example, it is possible to get colour-based ages
in excess of 9 Gyr in a galaxy which is only 4 Gyr old.
The important message is that the age
derived using this technique is not to be taken as a literal stellar 
population age; it is simply a way of parameterising how the 
recent SFR compares to the SFR several Gyr ago.

We can see from Fig.\ \ref{fig:c4age}, by comparison 
with the analagous figures for the closed box, infall and 
outflow models (Figs.\ \ref{fig:c4sc}, \ref{fig:c4inf} and \ref{fig:c4out}
respectively) that mass-dependent 
formation epoch variations between galaxies
affects their ages and metallicities
in ways similar to mass-dependent infall and outflow.  In particular, 
formation epoch differences generate magnitude--age and 
magnitude--metallicity correlations that the fiducial model lacks.

\subsection{Summary} \label{c4evosum}

To summarise:
\begin{itemize}
\item A local gas surface density-dependent Schmidt SFL
	describes many of the age and metallicity trends from 
	BdJ surprisingly well.  However, the slope 
	of the age--central surface brightness correlation and 
	the age gradients are underpredicted by the fiducial model.
\item If the decay timescale of infall is mass-dependent, it 
        can `imprint' mass-dependency
        on the galaxy ages without significantly affecting their 
	metallicities.
\item If the infall timescale depends on radius, an age
	gradient can be generated without significantly
	affecting the metallicity gradient.
\item If outflow is mass-dependent, it can `imprint' a metallicity--mass
	correlation without significantly affecting the ages of the model 
	galaxies.
\item Variations in galaxy formation epoch do not invalidate our conclusion
	that a local surface density-dependent SFL describes
	the data reasonably well.  However, mass-dependent variations
	imprint an age--magnitude and metallicity--magnitude relation
	even if no infall or outflow occurs.
\end{itemize}
In this way, the lack of mass dependence in the fiducial model
may be provided {\it either} by mass dependence in the formation epochs 
of galaxies (where less massive galaxies are younger), {\it or}
through a combination of mass dependent infall (where less massive
galaxies have longer infall timescales) and outflow (where less massive 
galaxies lose a larger fraction of their newly-synthesized metal content).
Furthermore, an `inside-out' formation scenario
may be able to reproduce the rather steep observed age
gradients within individual galaxies.

We are therefore left in an uncomfortable situation: is there any
way that we can differentiate between the effects of formation epoch variations
versus infall and outflow?  One possible approach is to study the 
{\it effective yield} of galaxies (i.e., the yield of a closed box model
which reproduces the observed metallicity of a galaxy at its 
observed gas fraction).  If the mass--metallicity
correlation is due to mass-dependent variations in formation epoch then the 
effective yield should be constant.  If, however, the mass--metallicity
correlation is generated by mass-dependent metal-enriched outflows
then the effective yield should strongly vary with galaxy mass, providing 
an observational test between the two possibilities.

\begin{figure}
\begin{center}
\psfig{figure=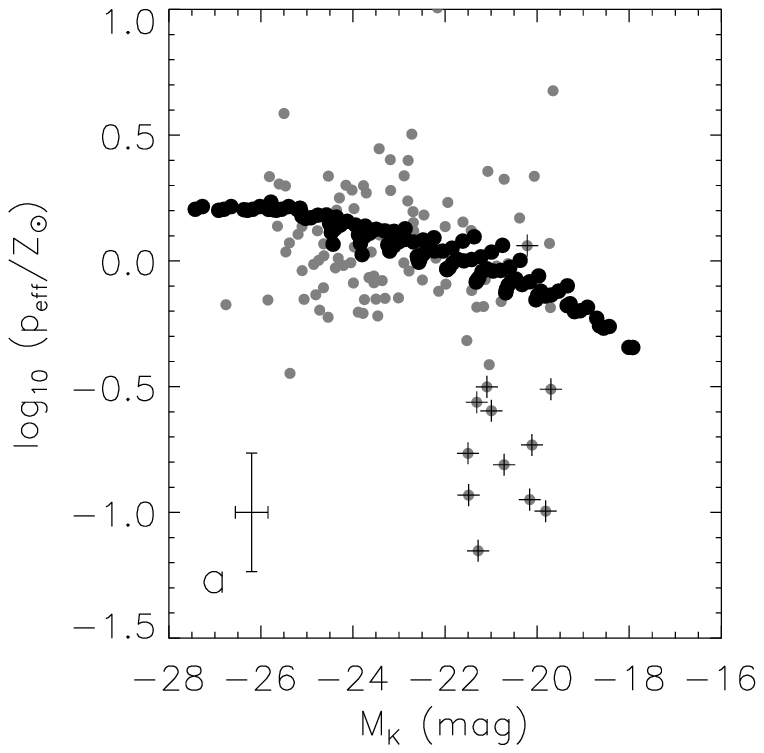,width=8cm}
\psfig{figure=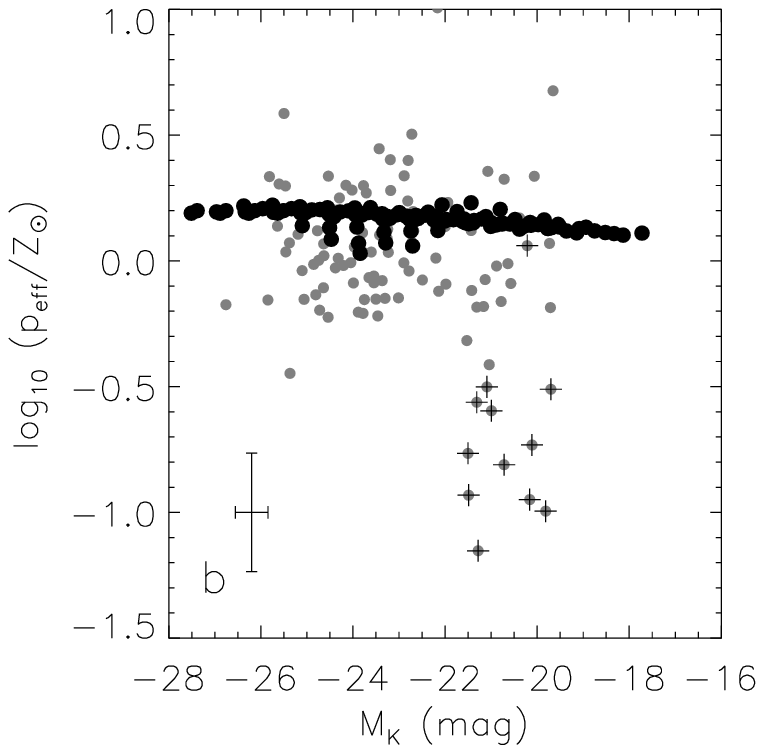,width=8cm}
\end{center}
\caption{Panel a: {\bf Outflow Model O.}, Panel b: {\bf Formation Epoch
Model E.}
The colour-based effective yield (see text for a brief
definition) of our sample galaxies against $K$ band magnitude 
for the galaxies from BdJ (grey circles) and the model (black 
circles).
Crosses denote data points with colour-based metallicities
less than 1/10 solar: these points may have substantially
underestimated metallicities due to model uncertainties
and thus substantially underestimated
effective yields (Bell et al.\ 2000; BdJ).
The error bar denotes the average error, taking account of 
metallicity and gas fraction uncertainties.
}
\label{fig:c4outeff}
\end{figure}

We show the result of this experiment in Fig.\ \ref{fig:c4outeff}
for the outflow and formation epoch models
respectively.  One can see that our naive expectation outlined
above is confirmed: the outflow model (panel a of Fig.\ \ref{fig:c4outeff}) has
a strongly mass-dependent effective yield, whereas the formation epoch
model has an almost constant colour-based effective yield
(panel b of Fig.\ \ref{fig:c4outeff}).  
The data do not clearly support either option:
the effective yield of the faint, metal-poor galaxies denoted
by crosses in Fig.\ \ref{fig:c4outeff}
may be substantially underestimated due to stellar
population model uncertainties at metallicities below 1/10 solar.
However, if trend indicated by these points is correct,  the
colour-based effective yields marginally support the outflow model, although 
discounting the formation epoch model on data with this much uncertainty
and scatter is clearly premature.

\begin{table}
\begin{center}
\caption{Model gradients}
  \label{tab:grads}
\begin{tabular}{lccc}
\hline
Model & \multicolumn{3}{c}{Gradients} \\
 & Age & Stellar met. & Gas met.\\
 & Gyr/$h$ & dex/$h$ & dex/$h$ \\
\hline
Fiducial & -0.47 & -0.13 & -0.18 \\
I & -0.89 & -0.16 & -0.22 \\
O & -0.48 & -0.13 & -0.18 \\
E & -0.52 & -0.14 & -0.19 \\
T & -1.02 & -0.16 & -0.22 \\
D & -0.85 & -0.14 & -0.37 \\
\hline
Observations & -0.78 & -0.13 & -0.20 \\
\hline
\end{tabular}
\end{center}
Model gradients are averages over all selected galaxies (see 
the selection criteria in Fig.\ \ref{fig:c4phys} and section
\ref{subsec:c4det}) and the observed gradients are averages over
the observational sample of BdJ (for age and stellar metallicity
gradients) and the sample of Garnett et al.\ 1997 for 
the gas metallicity gradients.
\end{table}

\section{Star formation laws} \label{c4sfl}

We have not yet considered changes in the SFL:
as we found that our results were quite sensitively affected by 
the SFL in the Schmidt law model, it is worth investigating if 
some of the shortcomings of the fiducial model can be 
alleviated by the use of an alternative SFL.

\subsection{Density threshold} \label{c4dens}

\begin{figure}
\begin{center}
\psfig{figure=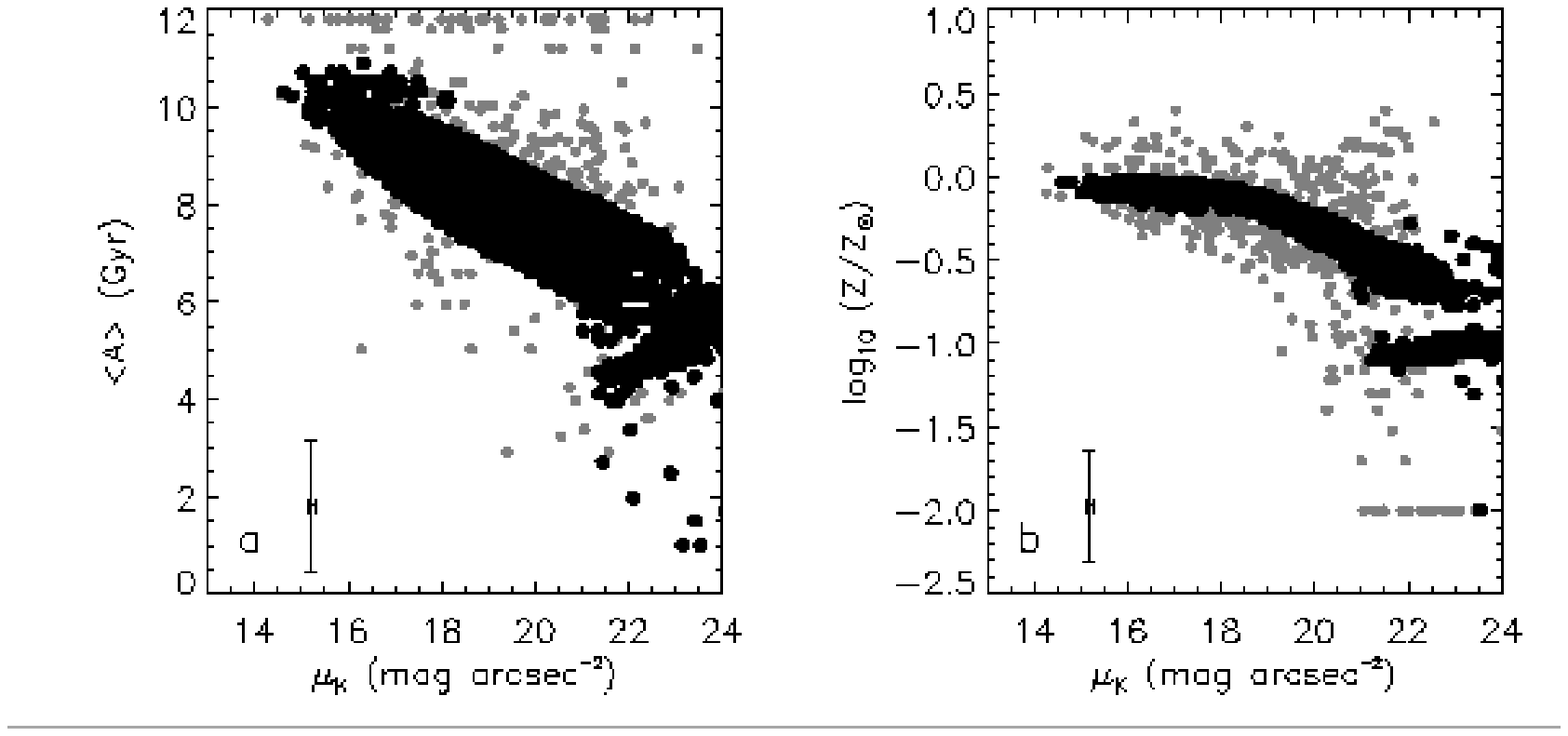,width=8cm}
\psfig{figure=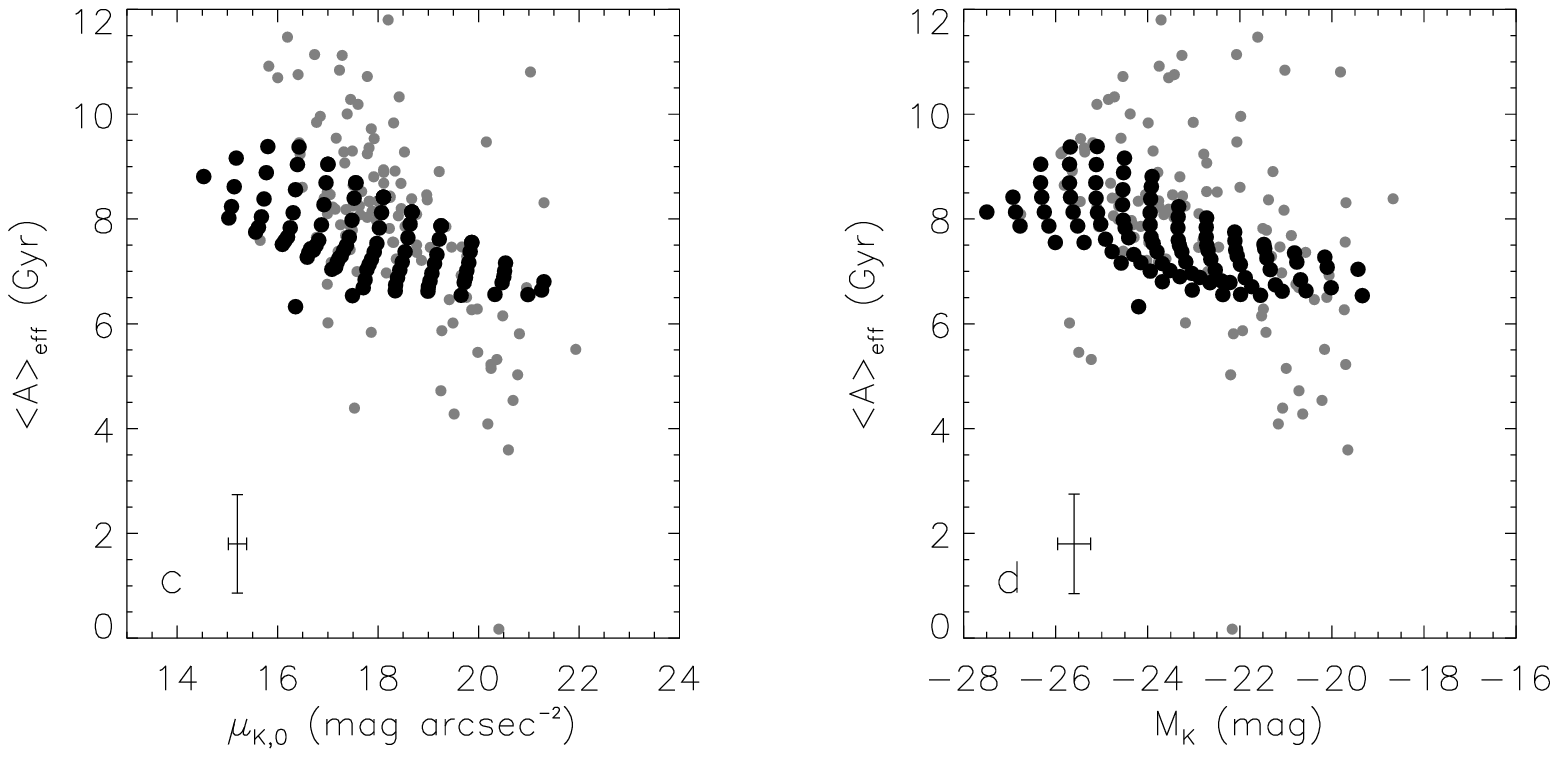,width=8cm}
\psfig{figure=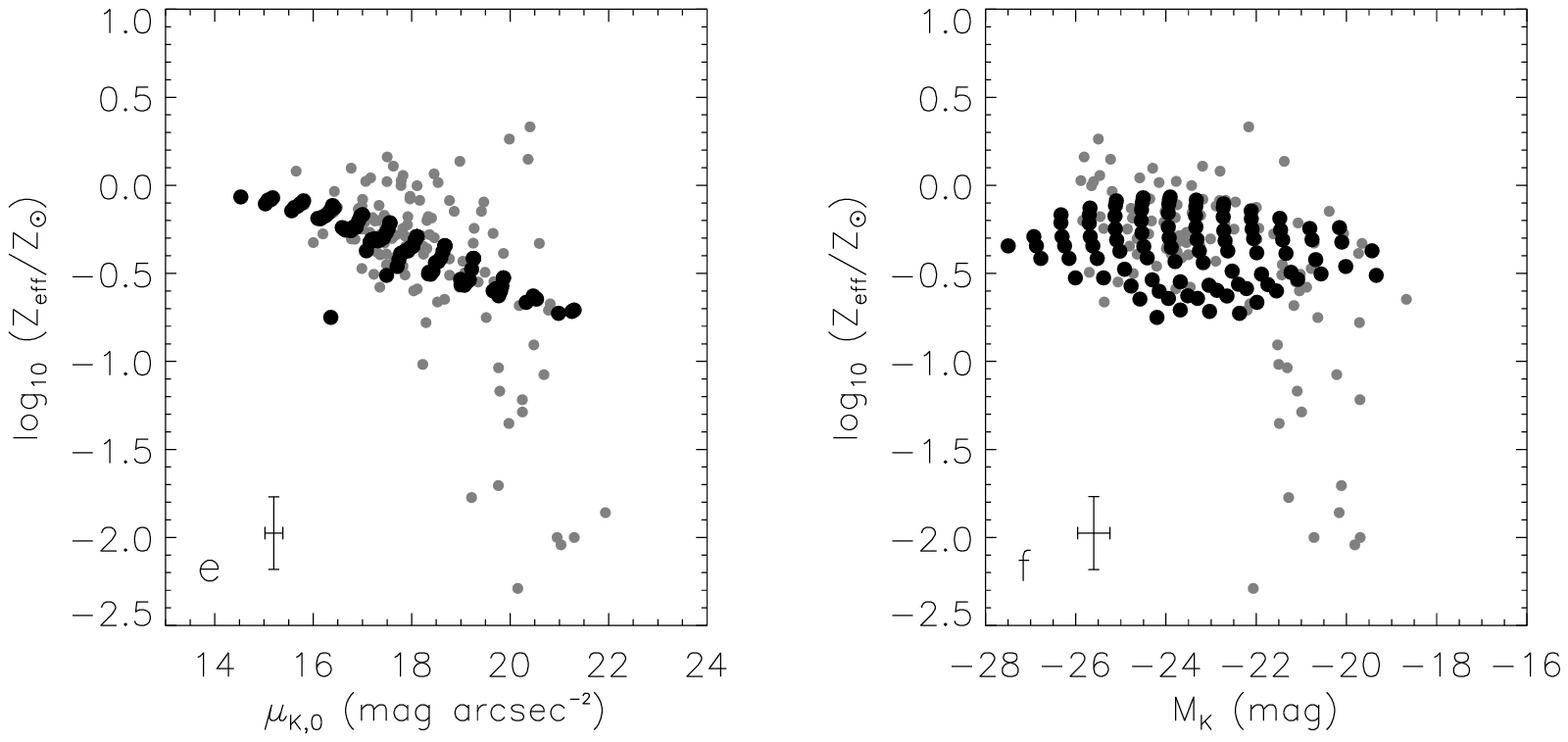,width=8cm}
\psfig{figure=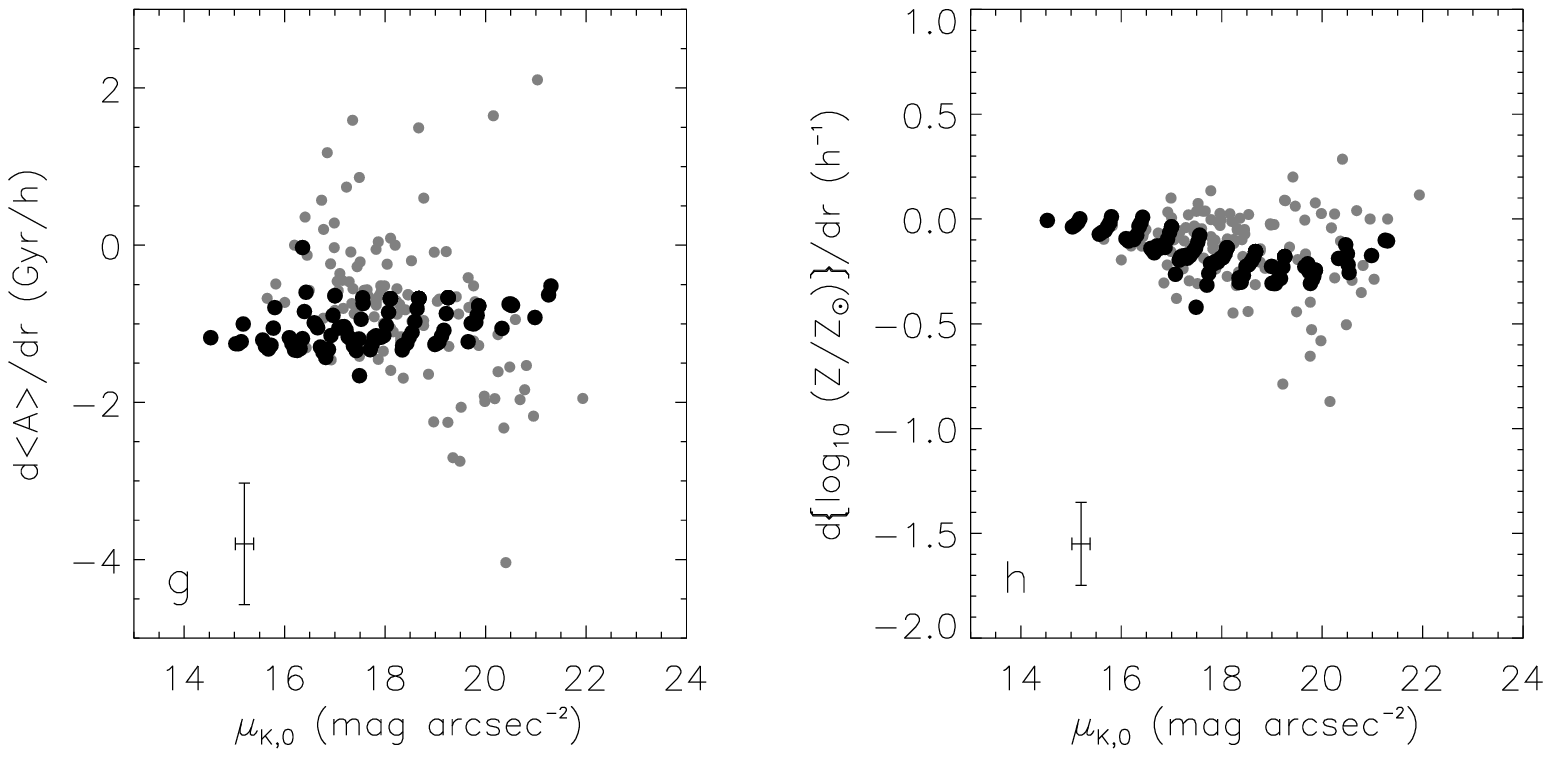,width=8cm}
\end{center}
\caption{{\bf Critical Density with Infall Model T: }
Panels and symbols are as in Fig.\ \protect\ref{fig:c4sc}}
\label{fig:c4k89i}
\end{figure}

One possible modification to the star formation law is the imposition 
of a cutoff critical density $\Sigma_c$ below which star formation 
cannot occur.  Kennicutt (1989) found that the radially-averaged
star formation rate (SFR) in a sample of 15 well-studied spiral galaxies
was well-described by a critical density model.
One physically motivated choice for that 
critical density is the maximum stable surface density
of a thin isothermal gas disc, given by (Toomre 1964; Kennicutt 1989):
\begin{equation}
\Sigma_c = \alpha \frac{\kappa c}{3.36G},
\end{equation}
where $\alpha$ is a dimensionless constant of order unity, $c$ is 
the velocity dispersion of the gas (which we take, as Kennicutt did, 
to be a constant
$c = 6$ km\,s$^{-1}$ for all galaxies) and $\kappa$ is the epicyclic
frequency
(Kennicutt 1989).
Kennicutt found that $\alpha \sim 0.67$ was a good fit to 
his data: stars typically did not form in regions where the density
was lower than the critical density, and formed according to a
Schmidt law with index $n \sim 1.3 \pm 0.3$ above the 
critical density threshold.

In order to estimate $\Sigma_c$ for our model galaxies, we must assume
a rotation curve: for simplicity, we adopt an adaptation of the 
mass-dependent `universal rotation
curve' from Persic \& Salucci (1991) (where
we adopt a baryonic mass of 
$1.5\times10^{11}\,{\rm M}_{\sun}$ for a $L^*$ galaxy 
with $B$ band absolute magnitude of $\sim -20.6$ 
assuming $H_0 = 65$ km\,s$^{-1}$\,Mpc$^{-1}$).
We use this rotation curve to 
determine the critical density in our model galaxies: we allow stars
to form according to a Schmidt law with $n = 1.8$ and $k = 0.012$ 
in gas denser than the critical density and do not allow stars to form at 
densities lower than the critical density.  {\it We do not use 
a critical density threshold in the central half scale length of the 
galaxy}:  apart from the undoubted influence of bars and bulges in the
central regions of galaxies (which we do not include in this model),
the universal rotation curve is undefined in the inner regions 
of a galaxy, and the critical density becomes very large in the innermost
regions of spiral galaxies due to strong differential rotation.
Thus, the behaviour of this star formation law at small radii is 
ill-constrained in this model: we, therefore, neglect the existence of 
a critical density at small radii and form stars according to 
a Schmidt law. 

In Fig.\ \ref{fig:c4k89i} we show the ages and metallicities
for a model with a critical density for star formation in which we
adopt infall (see section \ref{c4infall}).  Comparison with Fig.\ 
\ref{fig:c4inf} (and inspection of Table \ref{tab:grads})
shows that the imposition of a critical
density does little to affect the colour-based ages or metallicities
of spiral galaxies.  Only regions with relatively low gas densities 
are affected significantly:  regions near the centre of 
galaxies where the gas is depleted by star formation may be
affected (although note that we do not impose the critical 
density within the central half scale length of our model 
galaxies), and the outermost regions where infall
only recently brought the gas density above
the critical density.  In these outer regions 
the ages can be somewhat younger than those
of the infall model without a critical density (compare
the colour-based ages of those regions with the lowest $K$ band
surface brightnesses in panel a of 
Figs.\ \ref{fig:c4k89i} and \ref{fig:c4inf}).

To summarise: the critical density threshold proposed by 
Kennicutt (1989) does little to affect the global correlations
between SFH and physical parameters predicted by models 
with a Schmidt SFL.  This is not to say that the existence of a
critical density threshold has no effects at all: it is merely
to state that the existence of a critical density threshold
for star formation mainly affects the spatial distribution 
of present day star formation and does little to affect
the star formation or chemical enrichment histories as probed
by our colour-based technique.  

\subsection{Dynamical time dependence} \label{c4dyn}

An alternative SFL was proposed by 
Kennicutt (1998; Ke98 hereafter) based on his analysis of the global 
SFRs of a large sample of spiral and 
starburst galaxies.  The SFR
in this case scales with the ratio of the 
gas density to the local dynamical timescale:
\begin{equation}
\psi = k \Sigma_{\rm gas}/\tau_{\rm dyn},
\end{equation}
where $\tau_{\rm dyn} = 6.16 R(\rm kpc)/V(\rm km\,s^{-1})$ Gyr
is the dynamical timescale, here defined
as the time taken to orbit the galaxy at a distance $R$.
In this picture, the SFR is related to both the gas supply 
and the timescale in which the gas can be brought together.
As we did for the critical density model, we use the mass-dependent
universal rotation curve of Persic \& Salucci (1991) to 
estimate the dynamical time as a function of radius in our model galaxies.
This SFL is similar in many ways to the radially-dependent 
SFLs proposed by Wyse \& Silk (1989) and allows us to 
explore how explicit radial dependence in the SFL affects how
we interpret the trends in SFH with galaxy parameters 
presented in BdJ. 

\begin{figure}
\begin{center}
\psfig{figure=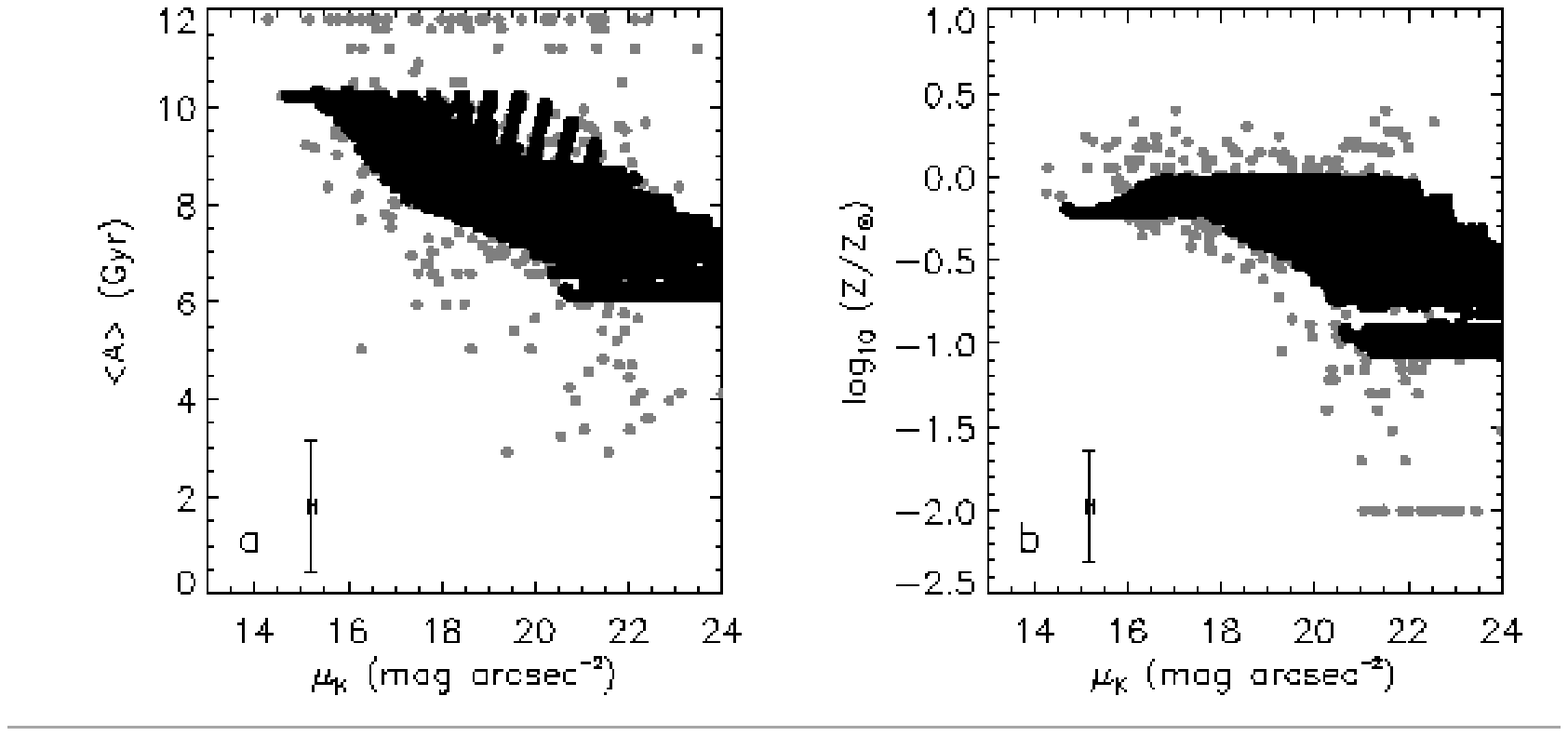,width=8cm}
\psfig{figure=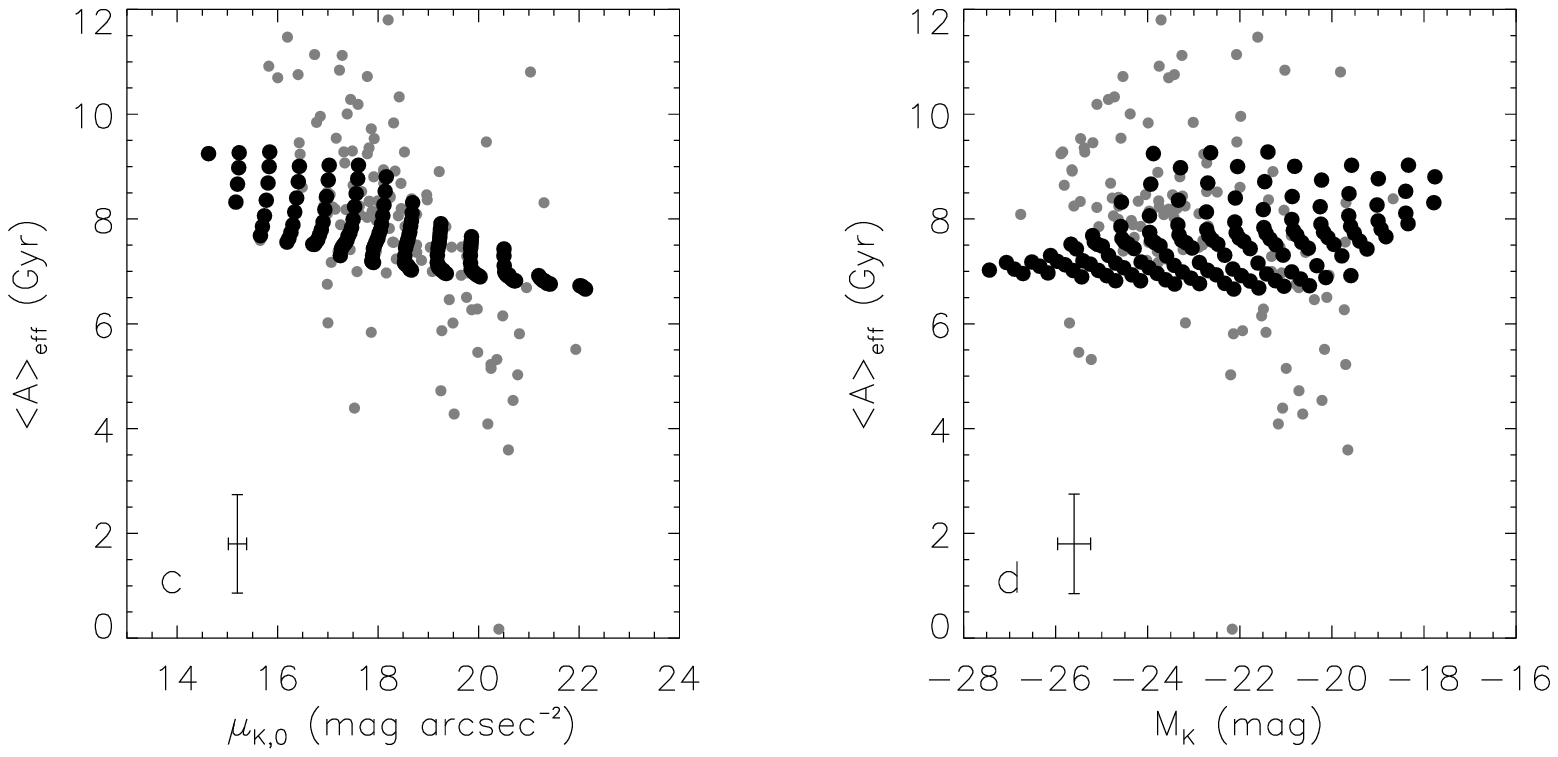,width=8cm}
\psfig{figure=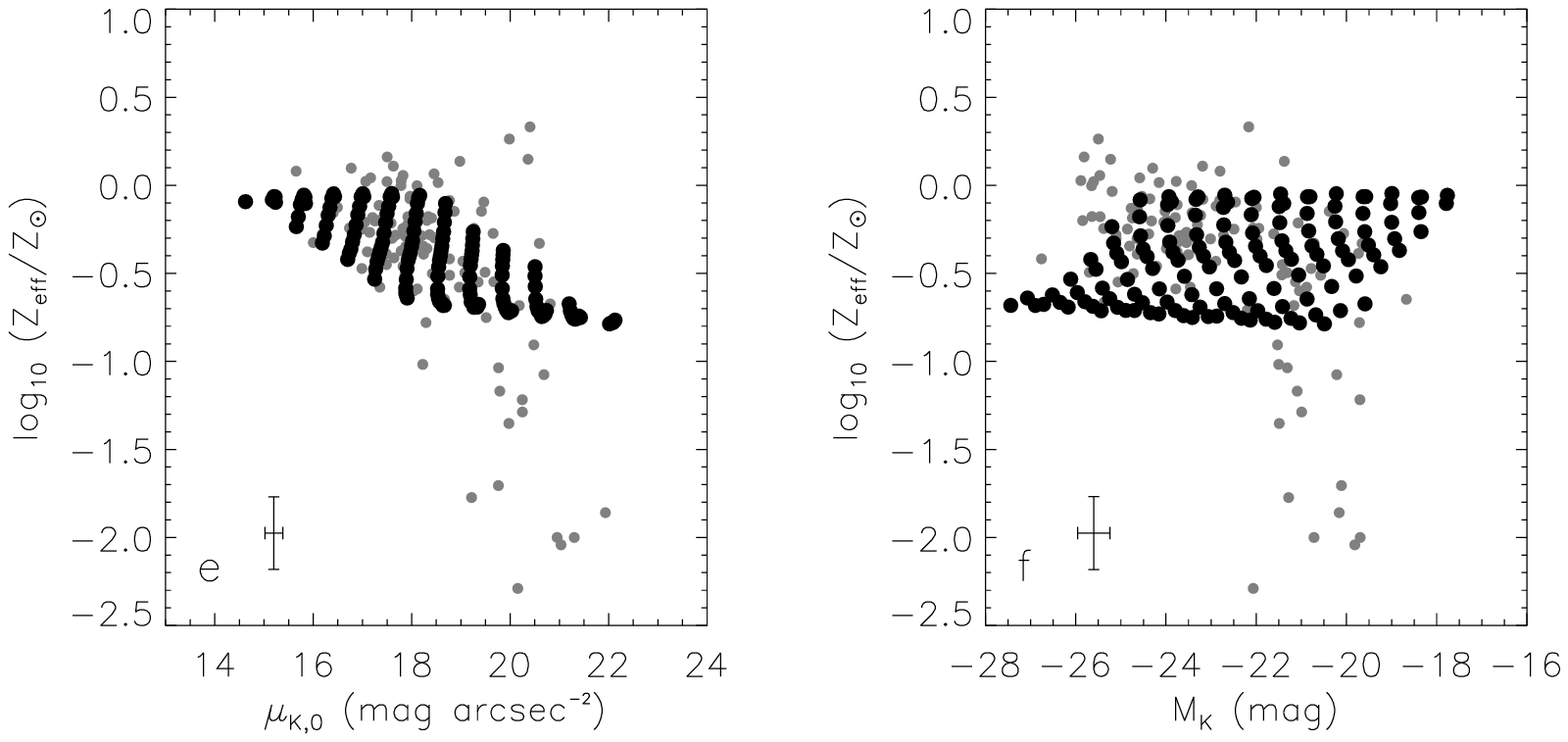,width=8cm}
\psfig{figure=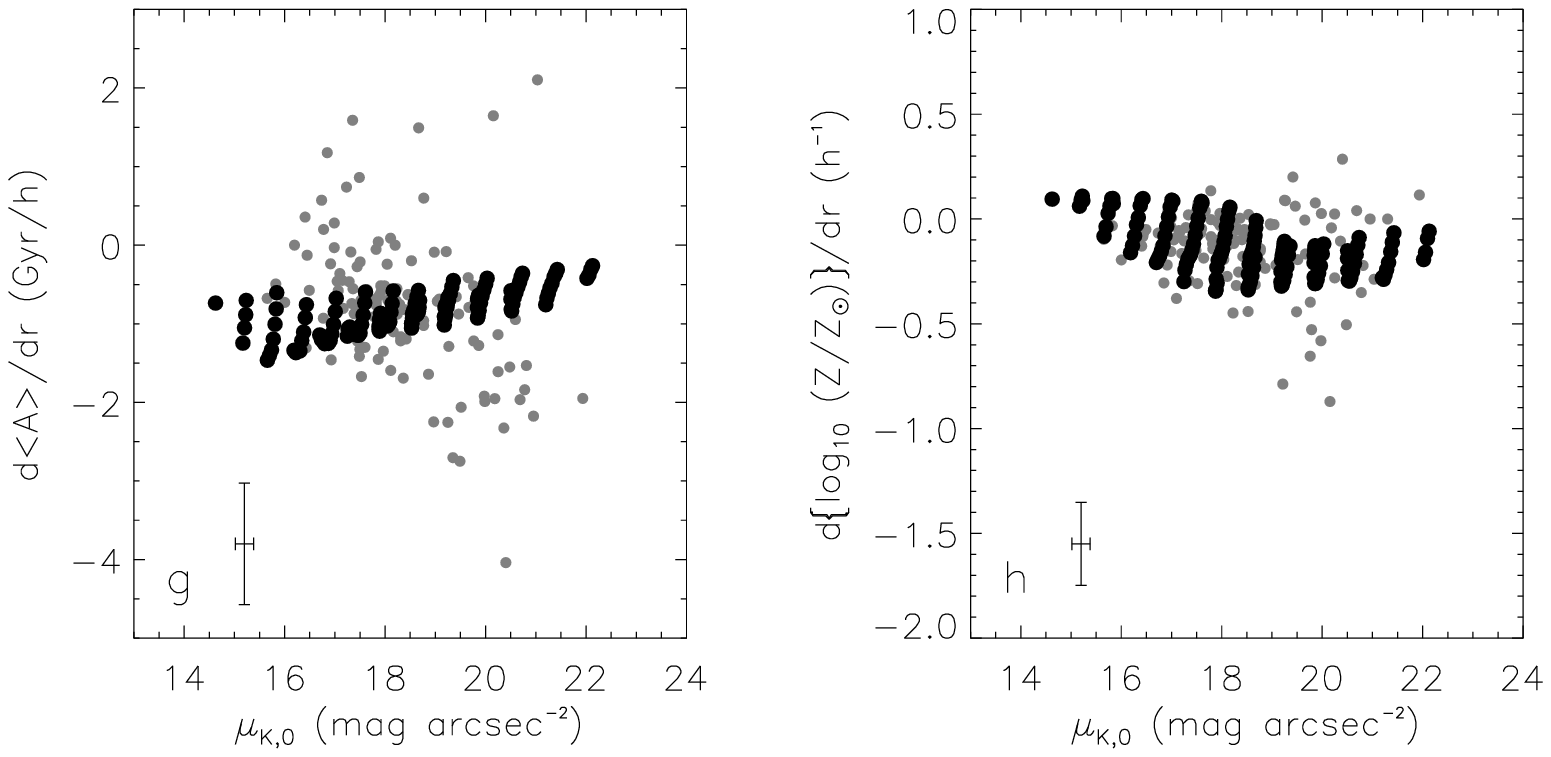,width=8cm}
\end{center}
\caption{{\bf Dynamical Time Model D: }
Panels and symbols are as in Fig.\ \protect\ref{fig:c4sc}}
\label{fig:c4k98}
\end{figure}

In Fig.\ \ref{fig:c4k98} we show the ages and metallicities
given by the dynamical time model (Model D) overplotted on the data
from BdJ.  The dynamical time dependence does three main things:
it introduces significant (but reasonable) scatter in the local ages and 
metallicities, it produces anticorrelations between 
age/metallicity and magnitude, and it produces steeper 
age and gas metallicity gradients.  The production of relatively
steep age and metallicity gradients is a success of this type 
of model.  Steep age gradients 
are expected from this model: in the limit of a flat rotation curve
the Ke98 SFL depends linearly on gas
density, and the star formation efficiency varies as $1/R$, leading
to larger star formation timescales at larger radii.
However, the observed {\it trend} in age gradient with 
surface brightness is not reproduced by the dynamical time model (or, indeed, 
the infall model): in both the dynamical time and infall model the age
gradients are steeper for higher surface brightness galaxies, which
is the opposite of the observed trend (Fig.\ \ref{fig:c4sc}).

Another more serious shortcoming of this model is the production of 
a `backwards' age--magnitude and metallicity--magnitude correlation.
This is at first sight counter-intuitive: brighter galaxies have larger
rotation velocities, which would increase the SFR at a given gas mass 
and physical radius.  However, brighter galaxies are also typically larger, 
therefore there is little change in the average ratio of radius to velocity, 
that is, there is little change in the typical dynamical time as 
a function of magnitude.  There is obviously a scatter in 
these properties, leading to a scatter in SFHs.  The 
`backwards' age/metallicity--magnitude correlation is generated
by the correlation between magnitude and surface brightness: 
small galaxies have older stellar populations in this model, and 
the smallest galaxies are also the faintest galaxies 
(see Fig.\ \ref{fig:c4phys}).  

To summarise: the dynamical time model, due to its 
explicitly radius-dependent SFR, generates steeper 
age and metallicity gradients, compared to the fiducial closed
box model.  However, the dynamical time, on average, does not
significantly depend on magnitude.  Coupled with the 
surface brightness--magnitude correlation, this implies
that the faintest galaxies appear, on average, older than 
the brightest galaxies, which is clearly at odds with the 
observations.  Therefore a dynamical time dependence, alone, 
is not favoured by our dataset.

\section{Discussion} \label{c4disc}

In contrast to the previous sections, where we tuned the models to improve
the match with the observational data, in this section we discuss three `blind'
applications of the models to literature data:
the models were not tuned to fit these observations
(solar cylinder data, global SFRs and gas metallicities), and they
represent a powerful test of the model's validity.
We also present a comparison of our models and the model
of Boissier \& Prantzos (2000).

\subsection{Properties of the Milky Way}

\begin{figure}
\begin{center}
\psfig{figure=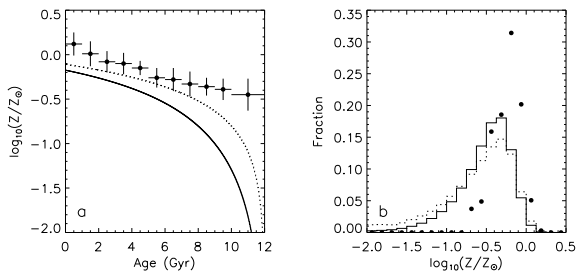,width=8cm}
\end{center}
\caption{
A comparison of the age-metallicity relation (panel a; 
points with error bars are from 
Rocha-Pinto et al.\ 2000a) and the oxygen metallicity distribution 
(panel b; solid dots are from Rocha-Pinto \& Maciel 1996
converted into oxygen abundance following Edvardsson et al.\ 1993)
of the solar circle with the model predictions for 
the infall model I (solid lines) and closed box fiducial
model (dotted lines). }
\label{fig:solar}
\end{figure}

Many chemical evolution models focus (sometimes almost
exclusively) on the Milky Way (e.g.\ Wyse \& Silk 1989; 
Steinmetz \& M\"{u}ller 1994; 
Prantzos \& Aubert 1995; 
Boissier \& Prantzos 1999; Portinari \& Chiosi 1999): 
only after the models have
been normalised to the Milky Way are the models extended
to external galaxies, if at all (e.g.\ Wyse \& Silk 1989;
Boissier \& Prantzos 2000; Prantzos \& Boissier 2000).  
We have 
approached the problem from the other direction: we use
the trends in galaxy properties with e.g.\ surface brightness
or magnitude to learn what processes might be at play in 
determining the star formation and chemical evolution histories
of galaxies.  Nonetheless, it is interesting to check our
models against observations of the ages and metallicities
of stars in the solar cylinder, to provide a consistency check
for our models.

In order to compare our models with the solar cylinder
age and metallicity distributions, it is necessary to
choose one model galaxy as a `Milky Way'.  We choose to
adopt a galaxy with a magnitude of around $M_K^* \sim -24$ 
and a $K$ band disc scale length of around 2.5 kpc in the closed
box model (this model galaxy has total mass of $10^{12}$ M$_{\sun}$
and a baryonic central surface density of 1000 M$_{\sun}$\,pc$^{-2}$).
In Fig.\ \ref{fig:solar} (panel a) 
we show the solar cylinder
age--metallicity relation 
(points and error bars; Rocha-Pinto et al.\ 2000a)
against the model predictions at a galactocentric
radius of 8.5 kpc for the fiducial model (dotted
lines) and the infall model I2 (solid lines).  Note that stars
from Rocha-Pinto et al.\ (2000a)
older than 10 Gyr and younger than 15 Gyr have been put in a 
single bin centred at 11$\pm$1 Gyr.
In panel b of Fig. \ref{fig:solar} we
show the metallicity (taken to be the oxygen abundance; determined
from their iron abundance by applying the trend in [O/Fe] with 
iron abundance from Edvardsson et al.\ 1993)
distribution of solar cylinder G stars taken from 
Rocha-Pinto \& Maciel (1996) against the model distributions
convolved with a gaussian of width 0.13 dex (to simulate the
intrinsic metallicity spread of stars at a
given age; Twarog 1980; Rocha-Pinto et al.\ 2000a).  

From Fig.\ \ref{fig:solar}, we see that the models
have some difficulties in reproducing the properties of the 
solar cylinder.  Panel b of Fig.\ \ref{fig:solar} clearly
demonstrates that both our closed box and infall models
have a significant G dwarf problem.  This was expected:
infall models require large amounts of late infall
to solve the G dwarf problem (e.g.\ Pagel 1998).  
Possible solutions of this problem include pre-enrichment
of the infalling gas 
from bulge and/or population III stars, more late infall, 
or some mixing of metals produced in outflows with 
the infalling gas.  
These additions are more complex processes that clearly need to 
be added to fine-tune the models.  However, since any of the models
could be tuned (in a few ways) to solve the galactic G dwarf problem,
this problem does not in itself help us choose between models, and 
so these processes are not considered here for simplicity.
The model age--metallicity relations
are also not a perfect match to the data: our model 
age--metallicity relations are offset to lower
metallicities, and show a steeper slope than the observed
age--metallicity relation.  While some of the slope mismatch 
may be due to the effects of observational errors (Rocha-Pinto et al.\ 2000a),
the mismatches in the age--metallicity relation are 
related to the G dwarf problem: the model 
metallicity is always lower than the observations, especially so
at early times. 

The Milky Way analogue 
model gas metallicity gradient is $-0.055$ dex\,kpc$^{-1}$,
which is broadly consistent with observational estimates of
between $-0.05$ and $-0.1$ dex\,kpc$^{-1}$ (e.g.\ Vilchez \& Esteban 1996;
Smartt \& Rolleston 1997; Pagel 1998).
We therefore conclude
that the models are a reasonable match to the observed ages and
metallicities of solar cylinder stars, modulo a G dwarf problem, 
and that our approach
of normalising the models on external galaxies has not 
seriously compromised the comparison with the Milky Way's properties.

\subsection{Comparison with the global star formation laws}

\begin{figure}
\begin{center}
\psfig{figure=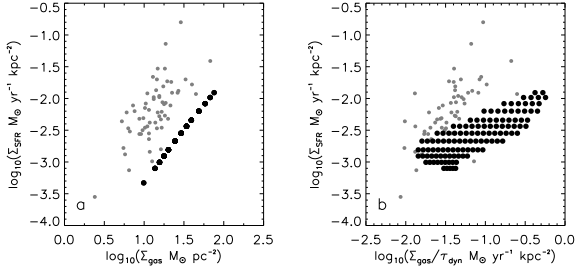,width=8cm}
\end{center}
\caption{{\bf Fiducial Model: }
A comparison of the gas density and
the star formation density interior to $R_{25}$ (left)
and the dynamical time SFR against
star formation density interior to $R_{25}$ (right).
The data (grey points) are spiral galaxies taken from Ke98 and
the SFRs from the fiducial model are denoted
by black circles.}
\label{fig:sfrfid}
\end{figure}

\begin{figure}
\begin{center}
\psfig{figure=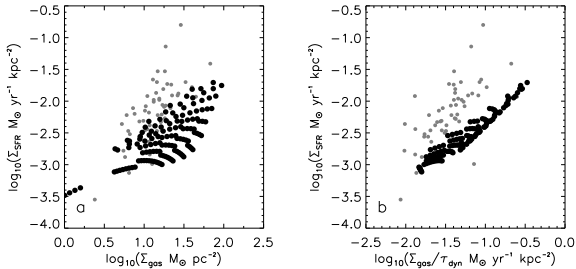,width=8cm}
\end{center}
\caption{{\bf Dynamical Time Model D: }
Panels and symbols are as in Fig.\ \protect\ref{fig:sfrfid}.}
\label{fig:sfrdyn}
\end{figure}

Another important test is the comparison of the SFLs
that we used in these models against observed SFRs.
Here, we compare our fiducial Schmidt law model and the dynamical 
time model with observations of the globally averaged SFRs,
gas densities and dynamical times of Ke98's subsample 
of spiral galaxies.  
To make the comparison fair, we construct gas densities and
SFRs averaged within the model galaxy's $R_{25}$ (i.e.\ 
interior to the 25 $B$ mag\,arcsec$^{-2}$ isophote), and we take
the dynamical time at $R_{25}$.

This comparison is presented in Figs. \ref{fig:sfrfid} and \ref{fig:sfrdyn}
for the fiducial model and dynamical time model respectively.
The left-hand panels show the globally averaged SFR as a function 
of the gas density, and the right-hand panels show the globally
averaged SFR as a function of the average gas density 
divided by the dynamical time.  The data are denoted by grey circles
and the models by black circles.  We can see that both models
provide a fair match to the slope of 
Ke98's observations, and encompass the 
range of SFRs and gas densities typical of spiral galaxies.
However, there is a significant
zero-point offset between the model and the observations.
A possible source of the zero-point offset is
the efficiency of the SFL $k$:  our model efficiencies
are somewhat lower than e.g.\ those efficiencies 
used by Boissier \& Prantzos (1999,2000).
(Somewhat higher star formation efficiencies would be required
to match the observations
if we assumed younger galaxy ages or postulated large amounts of 
late gas infall as Boissier \& Prantzos assume.)
A higher efficiency
raises the SFR at a given gas density, providing a better match 
to the observations.  However, there are other possibilities:
it is possible that there are aperture mismatches or discrepancies 
between the observational and model SFR calibrations (e.g.\ in the 
conversion of the \ha flux to SFR).  

\subsection{Comparison with gas metallicities}

\begin{figure}
\begin{center}
\psfig{figure=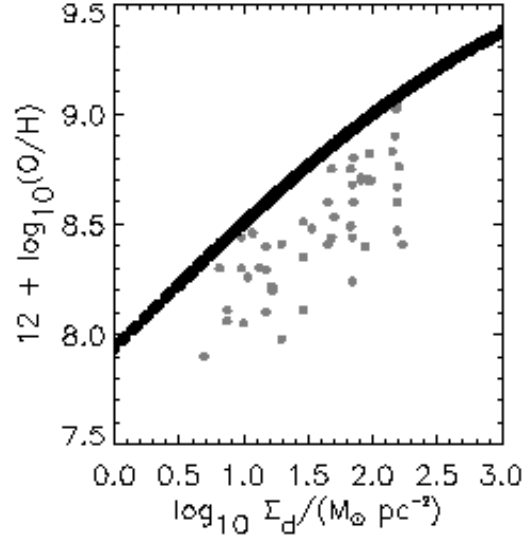,width=8cm}
\end{center}
\caption{{\bf Fiducial Model:  }
Comparison of trends in local gas oxygen abundance with baryonic
surface density with the fiducial model predictions.
The data are taken from Fig. 12  of Garnett et al.\
1997 (grey dots; for M33 and NGC 2403 with $B$ absolute 
magnitudes $\sim -18.5$), and the fiducial model predictions are overplotted
as black circles.}
\label{fig:gasmet}
\end{figure}

Earlier, we compared the models to the colour-based stellar ages
and metallicities of the sample of galaxies from BdJ. 
However, those ages and metallicities are subject to systematic
errors from e.g.\ model uncertainties and dust reddening.  For this
reason, it is important to check the models against the {\it gas
metallicities} derived from \hii region spectroscopy of spiral
galaxies: these metallicities are subject to a completely 
independent set of uncertainties, and as such offer a chance
to check our models.

We have chosen to compare our model gas metallicities with 
data from Garnett et al.\ (1997) in Fig.\ \ref{fig:gasmet},
where we show the surface density--local gas metallicity
correlation for M33 and NGC 2403 from their Fig.\ 12.  
Data from Garnett et al.\ (1997) are denoted by grey dots and
the fiducial model output is shown as black circles.
The model metallicity--surface
density correlation has the right slope but an offset zero point.

That the closed box fiducial model reproduces the right slope
of the metallicity--surface density correlation disagrees
with the 
conclusions of Phillipps \& Edmunds (1991), who concluded that observed
metallicity gradients exceed those achievable with a Schmidt SFL and 
closed box chemical evolution.  
However, inspection of their Figs.\ 1 and 2
suggests a good agreement between their simple Schmidt law model 
and the observations, indicating that they may have overestimated
the size of the observational trend between metallicity and surface
density when compared to the Schmidt law model.  

The offset zero point is a well-known problem:  when simple
models are used to infer a yield from observations of the 
oxygen abundances of spiral galaxies, the yield is estimated to be 
around 1/3 solar (e.g.\ Garnett et al.\ 1997; Pagel 1998).  
However, we require a yield of around or somewhat above solar to 
match the observations: this discrepancy amounts to an offset
of around 0.5 dex, which is of the order of the offset between the 
average metallicity of the models and the observations.  
This discrepancy between the yield estimates indicates
a mismatch between the colour-based and gas metallicities, 
although it is as yet uncertain whether this is a deficiency in
the calibration of the colour-based metallicities, 
gas metallicities, or even both.  However,
uncertainties in the stellar mass-to-light
ratios used to derive the surface densities will propagate
into this comparison.  In addition, M33 and NGC 2403
are relatively faint (M$_B \sim -18.5$), and because of the
metallicity--magnitude correlation they are expected to have
a somewhat lower metallicity than that of the fiducial model.

We do not show the comparison with their global gas metallicities
here: the results are consistent with the overall yield
offset between the gas and stellar metallicities outlined above
and, again, the well-known gas metallicity--magnitude correlation
(e.g.\ Skillman et al.\ 1989; Garnett et al.\ 1997)
can only be adequately reproduced by the outflow or formation epoch
model (as we found
for the stellar metallicity--magnitude correlation).
The average gas metallicity gradients of the model 
galaxies are presented in Table \ref{tab:grads}: the gas metallicity
gradient per disc scale length
of most models (with the exception of the dynamical time model)
is within 10 per cent of the observational average, indicating 
that gas metallicity gradients have relatively little power
to discriminate between the models.

\subsection{Comparison with Boissier \& Prantzos}

Recently, Boissier \& Prantzos discussed the properties
of a specific, comprehensive disc galaxy spectro-photometric
chemical evolution model in a trio of papers (Boissier
\& Prantzos 1999,2000; Prantzos \& Boissier 2000).
They constructed a specific spectro-photometric chemical
evolution model that reproduced many observational constraints, 
and explored why these observational constraints were reproduced
by their model.  We have taken a complementary approach,
where we have explored a wide range of physical processes to 
understand the effects of each process on the ages and metallicities
of spiral galaxies.  In this section, we compare the results of 
our modelling and theirs with a two-fold aim: to check that 
our studies give consistent results, and to gain some insight
into how robust the conclusions of Boissier \& Prantzos
are likely to be if any of the modelling details were changed.

Their model used a combined Schmidt and dynamical time 
SFL: $\psi = k\Sigma_{\rm gas}^{1.5}/\tau_{dyn}$, including 
surface density and mass-dependent infall, but, like our
models, they did not include gas flows.  They used a more 
sophisticated approach in dealing with the chemical evolution
than adopted in this work
(they did not use the IRA, but treated the full chemical 
evolution of the galaxy explicitly).  They used halo circular
velocities and spin parameters to parameterise their models
but did not use the information about halo formation to fully
specify the ages and infall histories of the discs they constructed
(c.f.\ Dalcanton et al.\ 1997; Mo, Mao \& White 1998).  Instead,
they assumed a constant galaxy formation epoch of 13.5 Gyr 
and tuned the surface
density and mass dependence in the infall history to provide as
good a match to the observations as possible.

Here, we focus on two issues: the origin of
age and metallicity gradients in spirals, and the nature
and origin of the mass--metallicity correlation.

One of the main conclusions of Prantzos \& Boissier
(2000) is that they required radial variation in both 
the infall timescale and the SFL to produce colour and
metallicity gradients that were large enough to agree with 
the observations.  However, we found earlier that explicit
radial dependence of {\it both} the SFL and infall history is
not required to reproduce stellar and gas metallicity gradients,
and that radial dependence in either one can produce a sufficient effect.

Another of the significant successes of the model from Boissier \&
Prantzos (2000) is the reproduction of the metallicity--magnitude
correlation.  However, at first sight, this is somewhat of a puzzle
as they do not include mass-dependent outflows or formation epoch differences
in their model: their metallicity--magnitude correlation is 
reproduced entirely by a mass-dependent infall timescale.
This puzzle is resolved by inspection of Fig.\ 3 of Boissier \&
Prantzos (2000):  it shows that at surface densities typical
of spiral discs, the infall prescription that they adopt 
{\it increases} with time for galaxies with halo circular
velocities $\la 150$ km\,s$^{-1}$.  This increasing infall rate 
mimics a variation in formation epoch.  
Therefore, we must modify our earlier conclusions:
the infall history at early and intermediate
times modifies primarily the colour-based ages of spirals; however,
large amounts of late infall affects both the colour-based ages and 
metallicities of spirals.  Furthermore, if the amount of late gas 
infall depends on galaxy mass, then a metallicity--magnitude
relation can be generated.  In their model, the metallicity--magnitude
relation is generated by differences in galaxy formation epoch (essentially) 
and so the prediction of a mass-independent effective yield
will hold for their model, just as it holds for our mass-dependent 
formation epoch model.

\section{Conclusions and prospects} \label{c4conc}

We have constructed a simple family of chemical evolution models 
with the aim of gaining some 
insight into the origins of many of the trends in SFH with 
galaxy parameters presented in 
BdJ.  The model is used to generate
colour-based ages and metallicities, which are directly comparable
with those derived in BdJ.  We generated a grid
of model galaxies and selected only those which lie in a 
pre-defined region of the $K$ band absolute magnitude--central
surface brightness plane.  

Using this model, we have found the following:
\begin{itemize}
\item A local gas surface density-dependent Schmidt SFL describes
many of the colour-based age and metallicity trends from BdJ
surprisingly well.  A model of this type does not explain 
the mass dependence in SFH required by BdJ
and significantly underpredicts the age gradient in 
spiral galaxies and the slope of the age--central surface 
brightness correlation.

\item The global properties of the fiducial model can be improved in 
either of two ways. 
  \begin{itemize}
  \item[(i)] A combination of mass-dependent infall and metal enriched
outflow imprint independent mass correlations 
on the galaxy colour-based ages and metallicities. Smaller galaxies
have a more extended period of inflow in this model, and lose a greater
fraction of their freshly-synthesized metals. 
  \item[(ii)] Galaxy formation epoch varies systematically with galaxy
mass. If less massive galaxies are younger (i.e.\ if they assembled the bulk of
their gas content at very late times) we explain the mass--metallicity 
and mass--age correlation without resorting to outflow.
  \end{itemize}

\item Regarding the radial variations within galaxies:
  \begin{itemize}
  \item[(i)] If the infall timescale varies with radius, an age
gradient can be generated. This has little or no effect on the metallicity
gradient.
  \item[(ii)] Alternatively, Kennicutt's (1998) SFL
(which involves both gas density and the dynamical time) produces
both strong age and gas metallicity gradients and 
a reasonable scatter in the local age/metallicity--surface 
brightness correlation.  The main shortcoming of this
model is a `backwards' age/metallicity--magnitude correlation.
\end{itemize}
\end{itemize}

One deliberate limitation of our empirical approach 
is that we do not incorporate our model galaxies into a detailed
cosmological context.  For example,
cold dark matter cosmologies make well-defined predictions about
the formation mechanisms and infall histories of galactic discs
that we ignore (e.g.\ Steinmetz \& M\"{u}ller 1994; Mo, Mao \&
White 1998; Somerville \& Primack 1999; Cole et al.\ 2000).
A more realistic treatment of the formation of our initial discs
would be desirable, but is sensitive to the poorly understood
details of angular momentum evolution in forming disc galaxies 
(e.g.\ Navarro \& Steinmetz 1997; 2000).  In this paper our philosophy
has been to determine which of our conclusions
seem the most robust to model details, and which may change
if our initial disc formation was made more realistic.
Our study gives an indication of which 
of our conclusions (and those of Boissier \& Prantzos) 
are robust: i.e.\ which physical processes
must operate in any galaxy formation scenario to reproduce the observations.
From the above discussion, we find the following.
\begin{itemize}
\item Gas surface density determines the SFR, although other factors
	(such as dynamical time or a critical density for star formation) may 
	also influence the SFR.
\item Radial dependence in the SFL and/or the infall history is
	favoured.
\item The infall of gas {\it either} varies strongly with galaxy mass,
        peaking at late times in low-mass
	galaxies (as parameterised by our formation epoch model, E, 
	or the model of Boissier \& Prantzos), 
	{\it or} infall varies more weakly with galaxy mass but operates
        in conjunction with a higher efficiency of metal-enriched
        outflows in low-mass systems (Models I and O). The data (as it
        stands) marginally favours the latter option (Fig.\
	\ref{fig:c4outeff}).
\end{itemize}

However, the last conclusion is strongly dependent on the treatment of 
very low metallicity galaxies for which our colour-based ages and
metallicities are highly uncertain. A combination of models I and O receives 
further support from studies of resolved stellar populations:
there is ample evidence for older stellar populations in 
local faint, gas-poor and metal-poor
dwarf Spheroidal galaxies.
These cannot fit into the mass-dependent
age scheme postulated above (e.g.\ Grebel 1998; 
Hurley-Keller, Mateo \& Nemec 1998).  Of course,
it is quite possible that very low-mass dwarf galaxies have a 
metallicity--magnitude correlation driven primarily by outflows, and 
more massive galaxies (such as spirals, which have managed to keep
the bulk of their gas content) have a metallicity--mass correlation 
driven primarily by differences in galaxy formation epoch.  The key observable
is the effective yield of galaxies: the effective yield 
gives insight into whether a galaxy is 
simply under-evolved and metal-poor (like, perhaps, low
surface brightness disc galaxies; de Blok et al.\ 1996; 
Bell et al.\ 2000) or has had the bulk of its metals removed in gas 
outflows (as appears likely for dwarf Spheroidals: e.g.\ Dekel \&
Silk 1986).  
What is clear is that more work, both on observational
and theoretical fronts, is required to fully elucidate the origin of 
the mass--metallicity correlation.

\section*{Acknowledgements}

We thank the anonymous referee for their helpful suggestions and comments,
and Helio Rocha-Pinto for communicating results
in advance of their publication.
We would like to thank Don Garnett for providing 
data in electronic form and for his comments on 
the manuscript.  We acknowledge useful
discussions with and input from 
Matthias Steinmetz, Roelof de Jong, Rob Kennicutt and Andrew Benson.
EFB acknowledges financial support from the Isle of 
Man Education Department, NASA grant NAG58426 and 
NSF grant AST9900789.

\end{document}